\documentclass[fleqn,usenatbib]{mnras}

\usepackage[T1]{fontenc}
\usepackage{url}
\DeclareRobustCommand{\VAN}[3]{#2}
\let\VANthebibliography\thebibliography
\def\thebibliography{\DeclareRobustCommand{\VAN}[3]{##3}\VANthebibliography}

\usepackage{graphicx}	
\usepackage{amsmath}	
\usepackage{amssymb}	
\usepackage{multirow}
\usepackage{booktabs}

\usepackage{ulem}

\title[]{Parameter Estimation of LAMOST Medium-resolution Stellar Spectra}

\author[Li et al.]{
Xiangru Li$^{1}$\thanks{E-mail: xiangru.li@gmail.com},
Xiaoyu Zhang$^{1}$,
Shengchun Xiong$^{1}$,
Yulong Zheng$^{1}$,
and Hui Li$^{1}$
\\
$^{1}$School of Computer Science, South China Normal University, No. 55 West of Yat-sen Avenue, Guangzhou 510631, China\\
}

\date{Accepted XXX. Received YYY; in original form ZZZ}

\pubyear{2022}

\begin{document}
\label{firstpage}
\pagerange{\pageref{firstpage}--\pageref{lastpage}}
\maketitle

\begin{abstract}
This paper investigates the problem of estimating three stellar atmospheric physical parameters and thirteen elemental abundances for medium-resolution spectra from Large Sky Area Multi-Object Fiber Spectroscopic Telescope (LAMOST).
Typical characteristics of these spectra are their huge
scale, wide range of spectral signal-to-noise ratios, and uneven distribution in parameter space. 
These characteristics lead to  unsatisfactory results on the spectra with low temperature, high temperature or low metallicity.
To this end, this paper proposes a Stellar Parameter Estimation method based on Multiple Regions (SPEMR) that effectively improves parameter estimation accuracy.
On the spectra with {S/N $\geq 10$}, the precisions are 47 K, 0.08 dex, 0.03 dex respectively for the estimations of ($T_{\rm eff}$, $\log \,g$ and $\rm [Fe/H]$), 0.03 dex to 0.06 dex for elements C, Mg, Al, Si, Ca, Mn and Ni, 0.07 dex to 0.13 dex for N, O, S, K and Ti, while that of Cr is 0.16 dex.
For the reference of astronomical science researchers and algorithm researchers, we released a catalog for 4.19 million medium-resolution spectra from the LAMOST DR8, experimental code, trained model, training data, and test data.

\end{abstract}




\section{Introduction}
\label{sec:intro}
In this paper, we study the estimation problem of stellar atmospheric parameters and element abundances from Large Sky Area Multi-Object Fiber Spectroscopic Telescope \citep[LAMOST;][]{cui2012, liu2015} medium-resolution stellar spectra.
LAMOST, also known as the Guo Shoujing Telescope, is a large optical band observation equipment.
It is the telescope with the highest spectral acquisition rate in the world and has provided a lot of precious spectral data for astronomical researchers.
Since October 2018, LAMOST started the second stage survey program (LAMOST \uppercase\expandafter{\romannumeral2}), which conducts both low- and medium-resolution spectroscopic surveys \citep{Wang_2019}.
The wavelength coverages of the LAMOST medium-resolution spectra are [4950, 5350] \text{\AA} and [6300, 6800] \text{\AA} \citep{Rui_2019}.
LAMOST DR8 released 5.53 million medium-resolution spectra, and the signal-to-noise ratio ($\rm S/N$) of 4.19 million spectra in them are greater than 10 \citep{Wang_2019}.

From the large amount of spectroscopic data obtained during the LAMOST survey project, stellar parameters and elemental abundances can be estimated \citep{liu2014} for huge number of stars. These parameters and elemental abundances can be used to infer the stars' properties and their evolutionary history \citep{recioblanco:hal-03857433}.
So far, researchers have proposed many methods to estimate the stellar parameters of LAMOST spectra.
In addition, the LAMOST survey project has its own Stellar Parameter Estimation Pipeline \citep[LAMOST stellar parameter pipeline, LASP;][]{luo2015, wu2011automatic}. 
The LASP works by minimizing the cardinality distance between the observed spectrum and theoretical spectra to find the best matching template and accordingly give the parameter estimate for the observed spectrum \citep{prugniel2001database}.
The limitation of this traditional method is that the model compuational complexity depends more on the grid that generates the theoretical spectra rather than on the problem complexity.
This results in relatively low computational efficiency and another limitation of this method is the high-quality requirements for the observed data.
However, the LAMOST observational spectral library is characterized by large amount of data and wide range of signal-to-noise ratios. This leads to a large room for improvement in the parameter estimation of LAMOST spectra.

\begin{figure*}
\centering
\includegraphics[scale=0.44]{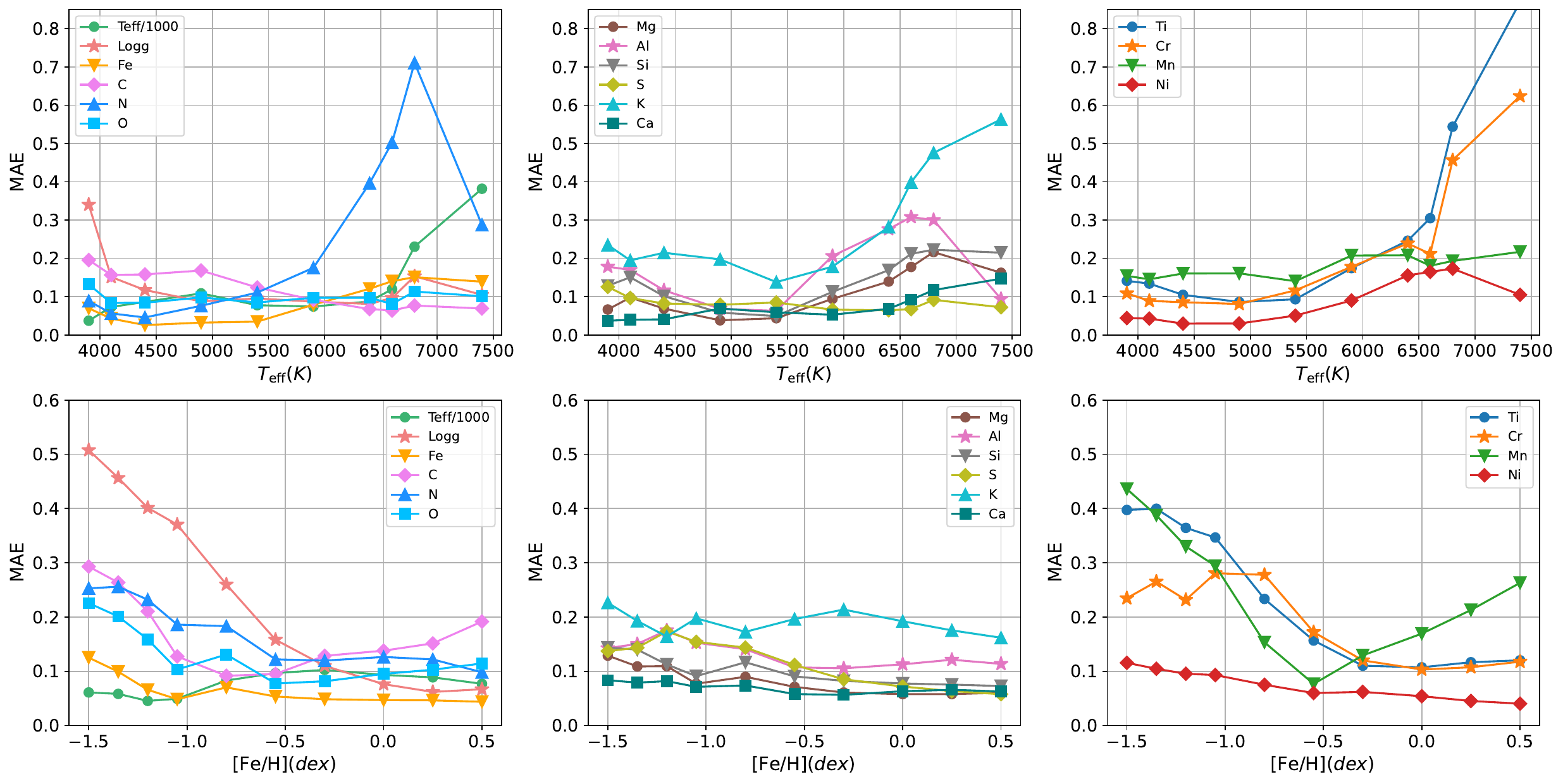}
\caption{The parameter estimation performance of RRNet: the performance obviously decreased on spectra with low temperature, high temperature or low metallicity.
Since RRNet is proposed for the problem of the parameter estimation for the medium-resolution spectra from LAMOST DR7, this experiment is carried out on 110,500 LAMOST DR7 spectra which have observations in the APOGEE DR17/ASPCAP catalog from common sources.
The performance of the parameter estimation is measured by the difference (Mean of Absolute Error, MAE) between the RRNet estimation and the reference result in the APOGEE DR17/ASPCAP catalog.
As the trained model and the complete model prediction results are not published by SPCANet, only the performance characteristics of RRNet are analyzed in this experiment.}
\label{fig:Teff_feh}
\end{figure*}

With the arrival of artificial intelligence and the big data era, researchers have tried to adopt deep learning methods to solve the problem of estimating stellar parameters from LAMOST medium-resolution spectra.
\cite{wang2020spcanet} proposed a residual-like network model (SPCANet) in 2020.
This model consists of three convolutional layers and three fully connected layers, and can accurately predict the stellar parameters and elemental abundances from LAMOST DR7 medium-resolution spectra.
In 2022, \cite{xiong2022model} developed a neural network model (RRNet) by combining several residual modules and some recurrent modules. 
The RRNet further improved the parameter estimation accuracy on the base of SPCANet.
However, the above two methods only effectively work on the spectra with a restricted parameter range. 
For example, the parameter estimation accuracy of SPCANet \citep{wang2020spcanet} on the spectra with $T_{\rm eff} > 6500$ K is significantly lower than that on the spectra with $T_{\rm eff} \in \rm [4000, 6500]$ K. 
Therefore, the SPCANet rejected the estimations for the spectra with $T_{\rm eff} > 6500$ K. 
RRNet \citep{xiong2022model} does not perform parameter estimation for spectra with $T_{\rm eff } < 4000$ K and  $T_{\rm eff} > 6500$ K.
In particular, the parameter estimation performance of RRNet is shown in Figure \ref{fig:Teff_feh}.
It is shown that the performance of the RRNet decreases apparently on the spectra with high temperature, low temperature, or low metallicity. The performance variation of the RRNet is closely related with the distribution characteristics of the observed spectra in the parameter space (More discussions can be found in Section \ref{sec:data}).

To deal with the above-mentioned problems, this paper proposes a Stellar Parameter Estimation method based on Multiple Regions (SPEMR) based on the distribution characteristics of LAMOST data in the parameter space.
This scheme significantly improves the estimation of parameters for the spectra with high temperature, low temperature, or low metallicity apart from its performance increasing on common type spectra. 
This paper is organized as follows:
Section \ref{sec:data} introduces the medium-resolution stellar spectra in LAMOST DR8, the reference set of this paper, and the scheme dividing reference set into different subsets according to the distribution characteristics.
Section \ref{sec:model} describes the principle of SPEMR.
The results of SPEMR on LAMOST DR8 are investigated in Section \ref{sec:result}.
Section \ref{sec:conclution} offers concluding remarks.

\section{Data}
\label{sec:data}

The model SPEMR proposed in this paper needs a reference set to learn the model parameters and to test model performance.
The reference set is established by cross-matching the LAMOST DR8 medium-resolution spectra with the APOGEE DR17/ASPCAP catalog.
The reference set consists of a series of samples, and each sample consists of an observed spectrum of an object and its reference label.
The reference spectra are obtained from the LAMOST DR8 medium-resolution spectral library, and the reference labels were the stellar physical parameters ($T_{\rm eff}$, $\log \,g$, [Fe/H]) and chemical abundances of 13 elements ([C/H], [N/H],
[O/H], [Mg/H], [Al/H], [Si/H], [S/H], [K/H], [Ca/H],
[Ti/H], [Cr/H], [Mn/H], [Ni/H]) from the APOGEE DR17/ASPCAP catalog.
It is worth noting that the reference sets provided by \cite{wang2020spcanet} and \cite{xiong2022model} are obtained by cross-matching the LAMOST DR7 medium-resolution spectral data with the APOGEE-\textit{Payne} catalog. 
While the reference set provided in this paper is based on the LAMOST DR8 medium-resolution spectra and APOGEE DR17/ASPCAP catalog.
The APOGEE DR17 catalog provides more reference labels and the corresponding labels with higher accuracy.
Therefore, we used the APOGEE DR17 catalog as the source of reference labels.
The reference set we obtained finally are more than twice as many as those of \cite{wang2020spcanet} and \cite{xiong2022model}.
This bigger reference set helps to build models with better accuracy of parameter estimation.
The typical characteristic of the reference set is that the data are exceedingly imbalanced in the parameter space (Figure \ref{fig:refer_dis}).
For example, in case of the effective temperature ($T_{\rm eff}$) higher than 6500 K, lower than 4000 K, or the metal abundance ($\rm [Fe/H]$) lower than -1.0 dex, the reference data are very sparse (Figure \ref{fig:Teff_feh}). 
This imbalance leads to a significant decrease in the accuracy of the parameter estimation models \citep[e.g.][]{wang2020spcanet, xiong2022model}.
To this end, this paper proposes a novel parameter estimation method based on multiple regions by dividing the parameter space into several sub-regions with different distrubution characteristics and accordingly dividing the reference set into three subsets.
More on the establishment of the reference set and its pre-processing procedures are described in the next two sub-chapters.

\subsection{Reference dataset based on common observational targets of APOGEE and LAMOST}
\label{subsec:refer_set}

\begin{figure*}
\centering
\includegraphics[scale=0.36]{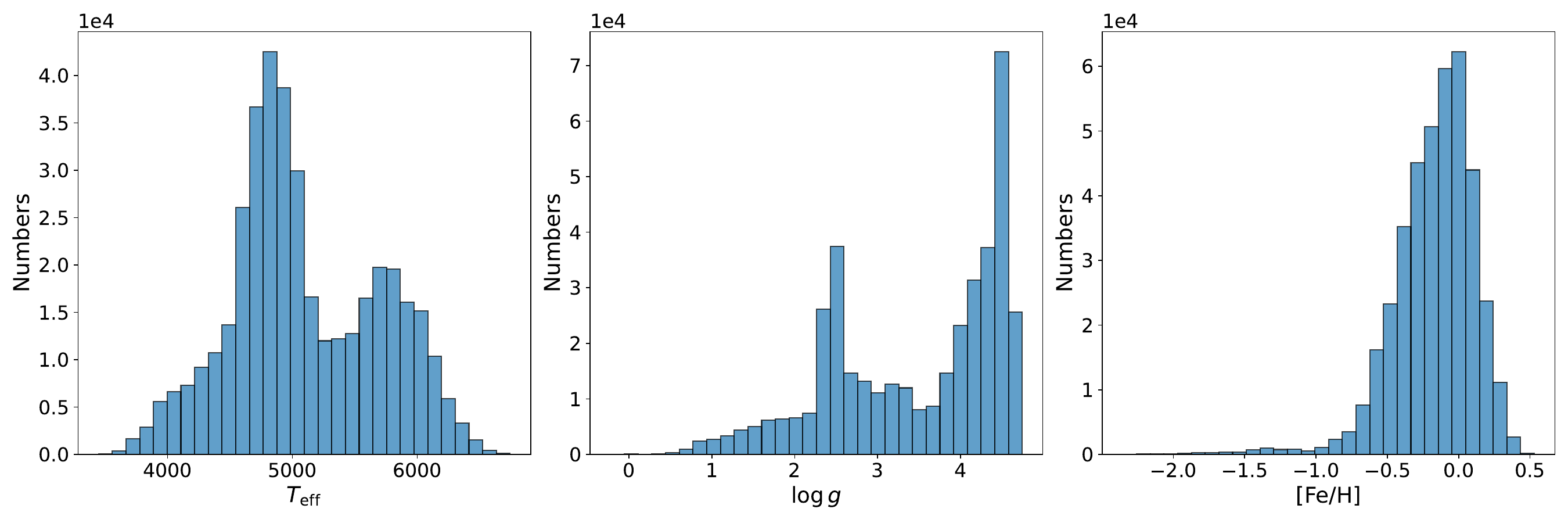}
\caption{The distribution histograms of the common sources between APOGEE DR17 catalog and LAMOST DR8 medium-resolution spectral library.}
\label{fig:APOGEE_DR17_distribution}
\end{figure*}

\begin{figure}
\centering
\includegraphics[scale=0.43]{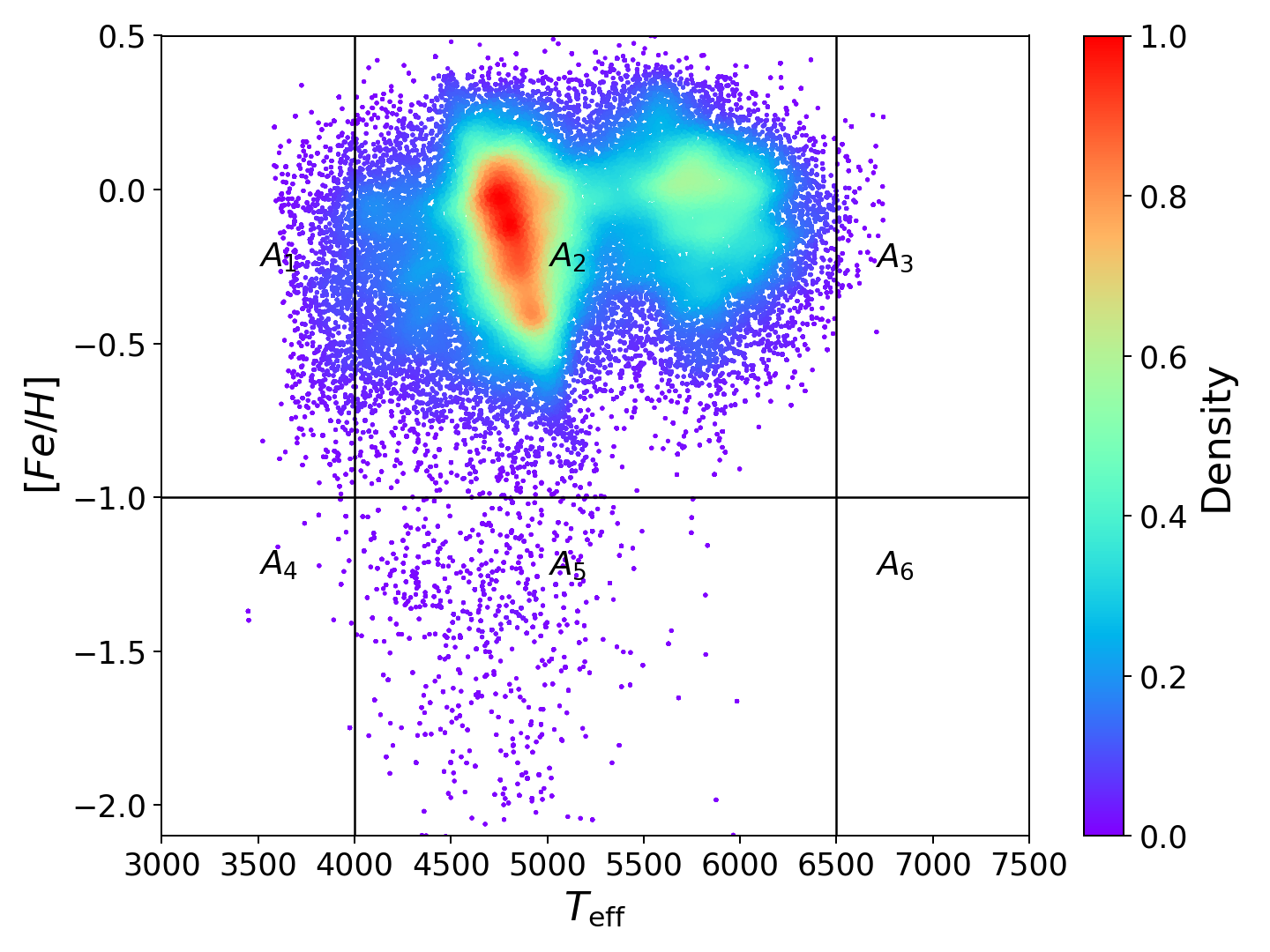}
\caption{The distribution of the overall reference dataset in the $T_{\rm eff}-\rm [Fe/H]$ space.
In this paper, the samples are divided into reference set 1 ($S_{1}$), reference set 2 ($S_{2}$) and reference set 3 ($S_{3}$) according to their distribution characteristics in the parameter space: $S_{1}=A_{2}$, $S_{2}=A_{1}\cup A_{3}\cup A_{4}\cup A_{6}$, $S_{3}=A_{4}\cup A_{5}\cup A_{6}$.}
\label{fig:refer_dis}
\end{figure}

\begin{figure}
\centering
\includegraphics[scale=0.43]{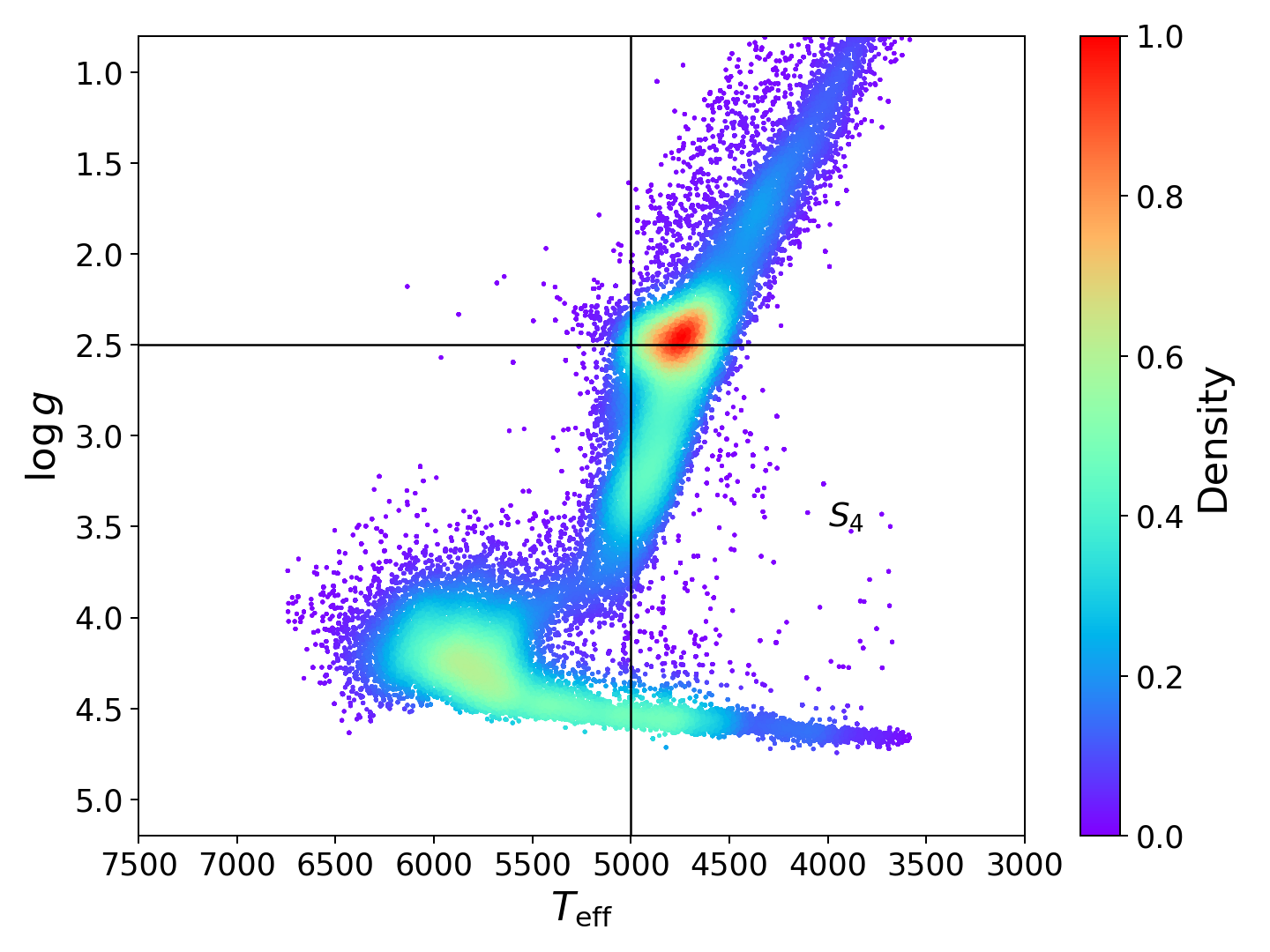}
\caption{The distribution of the overall reference dataset in the $T_{\rm eff}-\log g$ space.}
\label{fig:refer_dis_Teff_logg}
\end{figure}

APOGEE \citep{majewski2017apache} is a medium-high resolution (R$\sim$22500) spectroscopic survey in the near-infrared band ([15000, 17000] \text{\AA}).
The APOGEE spectra were obtained using the Sloan telescope at Apache Point Observatory in New Mexico City, USA.
The APOGEE Stellar Parameters and Chemical Abundances Pipeline \citep[ASPCAP;][]{perez2016} obtained stellar parameters and elemental abundances for most of the spectra by comparing the observed spectra with the theoretical spectral library using cardinal distance.
The APOGEE DR17 catalog publishes the stellar parameters ($T_{\mathrm{eff}}$, log $g$, [Fe/H]) and 20 elemental abundances for 475,144 stars.
The ranges of the stellar atmospheric parameters in the APOGEE DR17 catalog are $ \rm [3500, 7000]$ K for $T_{\rm eff}$, $\rm [-0.5, 5]$ dex for $\log \,g$, and $\rm [-2.0, 0.5]$ dex for $\rm [Fe/H]$.
The accuracies of the three parameters are 17 K, 0.03 dex and 0.009 dex, respectively.

In this paper, we used the same method as \cite{wang2020spcanet} and \cite{xiong2022model} to obtain the reference dataset.
We cross-matched the LAMOST DR8 medium-resolution spectra with the APOGEE DR17 catalog and obtained 75,316 common observational targets. 
There are 358,416 observed spectra in LAMOST medium-resolution spectral library from these common targets.
It is worth noting that some LAMOST spectra are affected by cosmic rays and other influences, which result in a large number of  outliers (bad pixels) in them. Therefore, the spectra with more than 100 outliers or more than 30 consecutive outliers are rejected. In addition, to ensure the reliability of the dataset, we only kept the spectral data with $\rm S/N \geq 10$ and $\rm quality\_flag = good$.
Finally, we obtained 73,773 common observational targets and 310,086 LAMOST DR8 medium-resolution spectra from these targets.
This dataset has over 100\% more data than the reference sets obtained by \cite{wang2020spcanet} and \cite{xiong2022model}. Figure \ref{fig:APOGEE_DR17_distribution} shows the distribution histograms of  the common sources between the APOGEE DR17 catalog and the LAMOST DR8 medium-resolution spectral library. 
It is shown that the data samples are sparse in the regions where $T_\mathrm{eff}$ > 6500 K, $T_\mathrm{eff}$ < 4000 K, log $g$ < 2.0 dex, and [Fe/H] < -0.5 dex.

To accurately predict the parameters for the spectra with high temperature, low temperature, or low metallicity, the obtained reference dataset is divided into three subsets according to the distribution characteristics of the samples in the $T_{\rm eff}-\rm[Fe/H]$ parameter space: reference set 1 ($S_{1}$), reference set 2 ($S_{2}$), and reference set 3 ($S_{3}$).
The three reference subsets are defined as shown in Figure \ref{fig:refer_dis}.
Reference set 1 ($S_{1}$) is used to further improve the parameter estimation accuracy on the spectra observed with high probability. 
Reference set 2 ($S_{2}$) is used to 
improve the parameter estimation accuracy on the spectra with high temperature or low temperature. And reference set 3 ($S_{3}$) is used to improve the parameter estimation accuracy on the spectra with low metallicity ($\rm[Fe/H]$). 
However, it is shown that the established model based on the above-mentioned three subsets performs unsatisfactory on the spectra of cool dwarf stars ($T_{\rm eff}$ < 4500 K and $\log \, g$ > 4.0 dex). Therefore, the fourth reference set,  $S_{4}$, is established (Figure \ref{fig:refer_dis_Teff_logg}). The thresholds $T_{\rm eff} < 5000$ K and $\log \, g > 2.5$ dex were chosen based on experimental performance.

\subsection{Data pre-processing}
\label{sebsec:data_process}

To facilitate machine learning model optimization, the reference spectra should be pre-processed before training the parameter estimation model. For example, wavelength correction, spectral resampling, spectral normalization, etc.
And the details of preprocessing procedure can be found in \cite{xiong2022model} and \cite{wang2020spcanet}.
After the above pre-processing procedures, the spectral data can be directly input into the SPEMR model for estimating the spectral parameters.

\section{Stellar Parameter Estimation based on Multiple Regions}
\label{sec:model}

In this paper, a novel method Stellar Parameter Estimation based on Multiple Regions (SPEMR) is proposed based on the distribution characteristics of LAMOST medium-resolution survey spectra in $T_{\rm eff}-\rm[Fe/H]$ and $T_{\rm eff}-\log \, g$ parameter space.
Since the parameter estimation for the spectra in each sub-region is implemented respectively based on the RRNet model, the proposed scheme can be specifically abbreviated as SPEMR (RRNet) in this paper.
The following sections will introduce the RRNet method, the motivation and principle of SPEMR model, and the method to obtain the final parameter estimation result for a spectrum based on SPEMR (RRNet), respectively.

\subsection{RRNet model}
\label{sec:model:RRNet}
Residual Recurrent Neural Network (RRNet) is a convolutional neural network whose main components are a recurrent learning module and a residual learning module \citep{xiong2022model}. 
RRNet model is proposed in the problem of parameter estimation of the medium-resolution spectrum of LAMOST.
Furtherly, compared with StarNet \citep[][]{fabbro2018application, bialek2020assessing} and SPCANet \citep{wang2020spcanet}, RRNet has some superiorities on accuracy and robustness.
Therefore, RRNet is chosen as the backbone network in the SPEMR model.

Compared to high-resolution spectroscopy, it is more challenging to discern some typical spectral line features in medium-resolution and low-resolution spectra.  
In these cases, it is necessary to design a parameter estimation algorithm with stronger sensitivity and detection capability for weak spectral features.
To this end, the RRNet model was proposed. 
In RRNet, the residual learning module enhances the sensitivity to spectral feature based on the driving power from parameter labels.

The super high spectral acquisition rate is a characteristic of the LAMOST survey, which helps to acquire a large-scale stellar spectral data set in a short period of time.
However, an accompanying problem is the large amount of noises in the observed spectra.
The recurrent learning module in the RRNet achieves cross-band information propagation and belief enhancement by mining the correlation between spectral features on different bands.
And this module can suppress the negative effects from noises in the spectra.
More information about RRNet can be found in \cite{xiong2022model}.

\subsection{Division of sub-regions and overall learning architecture}
\label{sec:model:SPEMR}

\begin{figure*}
\centering
\includegraphics[scale=0.63]{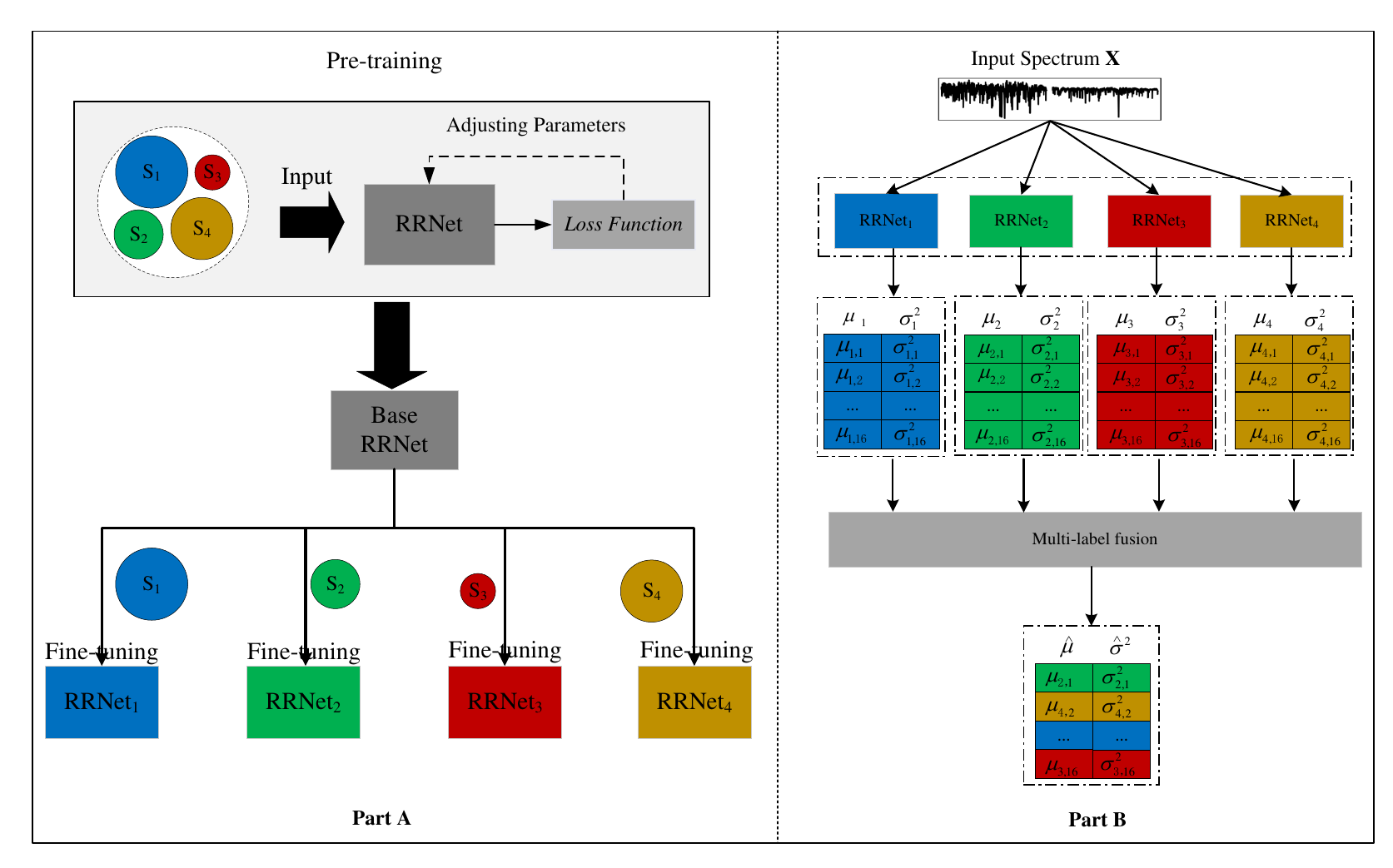}
\caption{The learning principle and prediction principle of the SPEMR model.
Part A shows the learning process of the SPEMR model, which consists of two phases: overall pre-training and personalized fine-tuning.
Part B shows the flow chart of SPEMR for parameter estimation on the spectra.
$S_1$, $S_2$, $S_3$ and $S_4$ denote the reference data used for training (Figure \ref{fig:refer_dis} and Figure \ref{fig:refer_dis_Teff_logg}).}
\label{fig:train_model}
\end{figure*}

A two-stage learning scheme is used in SPEMR to improve the accuracy of parameter estimation both on high-frequency-observed-type spectra and on spectra with low temperature, high temperature, or low metallicity. The two learning stages are an overall pre-training and a personalized fine-tuning (Part A in Figure \ref{fig:train_model}).
In the first stage, RRNet is trained by the reference spectra over the entire parameter space to obtain a common knowledge of the parameter estimation problem. 
In the second stage, four reference subsets $S_1$, $S_2$, $S_3$ and $S_4$ (Figure \ref{fig:refer_dis} and Figure \ref{fig:refer_dis_Teff_logg}) are independently used for further personalized optimizing the pre-trained model and four models $RRNet_1$, $RRNet_2$, $RRNet_3$ and $RRNet_4$ are obtained.  This fine-tuning learning allows the model to better handle specific types of spectral parameter estimation problems.
Four fine-tuned models $RRNet_1$, $RRNet_2$, $RRNet_3$ and $RRNet_4$ are fused into the final SPEMR parameter estimation model. 
More information about the two learning stages of SPEMR is presented below.

\subsubsection{Overall pre-training}
\label{sec:pre-training}

\begin{table*}
\centering
\caption{Hyperparameter evaluation. The $N_r$ and $N_s$ are two hyperparameters in RRNet.  
Experimental results are calculated on the validation set using the mean absolute error (MAE).
To simplify the computation complexity, the evaluations of $N_s$=5, 20, 40 and 60 are conducted with $N_r$ = 3, and evaluations of $N_r$= 1, 2, 3, 4 and 5 are conducted with $N_s$=5.}
\begin{tabular}{lcccc|cccc}
\hline
Labels
        & $N_r$=1   & $N_r$=2            & $N_r$=3             & $N_r$=4            & $N_s$=5    & $N_s$=20           & $N_s$=40            & $N_s$=60           \\ \hline
    
   $T_{\rm eff}$  &42.6130	    &\textbf{40.0473}	&41.7769	&40.1840  &41.7769  &40.3235    &\textbf{39.8680}  &40.3373  \\
    $\log \,g$    &0.0807	    &0.0737	    &0.0744	    &\textbf{0.0725}	  &0.0744	&0.0719	    &0.0718	  &\textbf{0.0716}\\
    {[}Fe/H{]}    &0.0286	    &\textbf{0.0276}	    &0.0282	    &0.0281	  &0.0282	&0.0279	    &\textbf{0.0278}	  &0.0280\\
    {[}C/H{]}    &0.0427	    &0.0417	    &\textbf{0.0413}	    &0.0418	  &\textbf{0.0413}	&0.0417	    &0.0419	  &0.0423\\
    {[}N/H{]}    &0.1030	    &0.1023	    &\textbf{0.1015}	    &0.1024	  &\textbf{0.1015}	&0.1028	    &0.1034	  &0.1040\\
    {[}O/H{]}    &0.0554	    &0.0552	    &\textbf{0.0551}	    &0.0552	  &\textbf{0.0551}	&0.0552	    &0.0554	  &0.0555\\
    {[}Mg/H{]}    &0.0327	    &0.0320	    &\textbf{0.0318}	    &0.0321	  &\textbf{0.0318}	&0.0322	    &0.0323	  &0.0327\\
    {[}Al/H{]}    &0.0472	    &0.0462	    &\textbf{0.0458}	    &0.0463	  &\textbf{0.0458}	&0.0466	    &0.0469	  &0.0473\\
    {[}Si/H{]}    &0.0339	    &0.0335	    &\textbf{0.0331}	    &0.0334	  &\textbf{0.0331}	&0.0337	    &0.0338	  &0.0340\\
    {[}S/H{]}    &0.0624	    &0.0622	    &\textbf{0.0621}	    &0.0623	  &\textbf{0.0621}	&0.0624	    &0.0623	  &0.0624\\
    {[}K/H{]}    &0.0638	    &0.0636	    &\textbf{0.0634}	    &0.0638	  &\textbf{0.0634}	&0.0639	    &0.0640	  &0.0642\\
    {[}Ca/H{]}    &0.0376	    &0.0370	    &\textbf{0.0367}	    &0.0369	  &\textbf{0.0367}	&0.0372	    &0.0372	  &0.0374\\
    {[}Ti/H{]}    &0.0919	    &0.0912	    &\textbf{0.0915}	    &0.0916	  &\textbf{0.0915}	&0.0918	    &0.0921	  &0.0924\\
    {[}Cr/H{]}    &0.1086	    &0.1081	    &\textbf{0.1079}	    &0.1076	  &\textbf{0.1079}	&0.1080	    &0.1082	  &0.1083\\
    {[}Mn/H{]}    &0.0440	    &0.0431	    &\textbf{0.0425}	    &0.0427	  &\textbf{0.0425}	&0.0429	    &0.0428	  &0.0431\\
    {[}Ni/H{]}    &0.0357	    &0.0351	    &\textbf{0.0348}	    &0.0349	  &\textbf{0.0348}	&0.0348	    &0.0349	  &0.0351\\
    
    \hline
    \end{tabular}
    \label{tab0}
\end{table*}

In the pre-training process, we randomly divide the overall reference set (see \ref{subsec:refer_set} section) into a training set, a validation set and a test set at the ratio of 7:1:2. 
The three data sets respectively consist of 217,379 spectra from 51,641 stars, 30,821 spectra from 7,377 stars, and 61,886 spectra from 14,755 stars.
The training set is used for learning the pre-trained model parameters, the validation set is used to determine the pre-trained model hyperparameters, and the test set is used to evaluate the performance of the  parameter estimation results.

To accurately estimate the probability density function \citep[PDF;][]{bialek2020assessing} of the estimated stellar parameters, 6 instances of the model are trained with different random initializations.
The mean {$\hat \mu(\mathbf{X})$} of the ensembling is determined by the average of the predicted means of these six models. 
The variance {$\hat \sigma^2_{pred}(\mathbf{X})$} of the ensembling is determined by the following equation:
\begin{equation}
    \label{equ:final_variance}
    \hat\sigma^2_{pred}(\mathbf{X}) = \frac{1}{6} \sum_{i=1}^6{(
    \sigma^2_{\theta_i}(\mathbf{X})+ \mu^2_{\theta_i}(\mathbf{X})
    ) 
    -\hat\mu^2(\mathbf{X})},
\end{equation}
where {$\theta_i$} is the parameters to be optimized for the i-th model, and {$\mu_{\theta_i}(\mathbf{X}), \sigma^2_{\theta_i}(\mathbf{X})$} are the mean and variance of the prediction of the i-th model, respectively.

In the RRNet model, a spectra is divided into $N_s$ sub-bands. 
The correlation and complementarity of spectral information between various bands are learned through the recurrent module.
There is another hyperparameter $N_r$ in the RRNet model, which indicates the number of residual blocks.
These two hyperparameters have an impact on the RRNet model performance. 
Therefore, we optimized them using the validation set.
In this paper, some experimental explorations are conducted on different configurations $N_r$ = 1, 2, 3, 4 and $N_s$ = 5, 20, 40, 60 (Table \ref{tab0}).
It is shown that the pre-trained model is the smallest error on the whole in case of $N_r$ = 3 and $N_s$ = 5.
Therefore, the $N_r$ of RRNet are set to 3 and the $N_s$ are set to 5 in the subsequent experiments. 
In addition, the number of training iterations and the learning rate are 30 and $10^{-4}$, which are consistent with RRNet \citep{xiong2022model}.

\subsubsection{Personalized fine-tuning}
\label{sec:fine-tuning}
Although the Base RRNet obtained in the pre-training stage already has some parameter estimation capabilities in the overall parameter space, the non-uniformity of the sample distribution (Fig. \ref{fig:refer_dis} and Fig. \ref{fig:refer_dis_Teff_logg}) leads to a significant room for improvement in each sub-region (Fig. \ref{fig:Teff_feh}).
Therefore, we fine-tuned the model in a targeted way for each sub-region separately.
Specifically, the spectra with $T_{\rm eff} \in \rm [4000, 6500]$ K and $\rm[Fe/H] \geq -1.0$ dex in the training set are used as training set 1 (the $S_1$ in Fig. \ref{fig:refer_dis}) to fine-tune the Base RRNet, and the corresponding parameter estimation model $RRNet_1$ is obtained. Furtherly, the spectra with $T_{\rm eff} < 4000$ K and $T_{\rm eff} >6500$ K in the training set are treated as training set 2 (the $S_2$ in Fig. \ref{fig:refer_dis}) to fine-tune the Base RRNet, and the corresponding parameter estimation model $RRNet_2$ is computed. 
The spectra with $\rm[Fe/H]<-1.0$ dex in the training set are considered as training set 3 (the $S_3$ in Fig. \ref{fig:refer_dis}) to fine-tune the Base RRNet, and the corresponding parameter estimation model $RRNet_3$ is obtained.
Finally, the spectra with $T\mathrm{eff}$ < 5000 K and $\log g$ > 2.5 dex in the training set are considered as training set 4 (the $S_4$ in Fig. \ref{fig:refer_dis_Teff_logg}) to fine-tune the Base RRNet, and the corresponding parameter estimation model $RRNet_4$ is obtained.

When fine-tuning each sub-model, we kept the parameters of the convolutional layer unchanged, and only re-optimized the parameters of the fully connected layer. 
In this way, the sub-model can converge earlier and has a relatively strong spectral feature extraction ability at the beginning.
In addition, the number of training iterations and the learning rate for each sub-model are respectively set to 10 and $10^{-5}$. This method also further accelerates the convergence of the sub-models.
After fine-tuning, we obtained four sub-models, $RRNet_1$, $RRNet_2$ and $RRNet_3$, $RRNet_3$ and $RRNet_4$.
The $RRNet_1$ is used to predict the stellar parameters for the stellar spectra observed with high probability. The $RRNet_2$ is used to predict the stellar parameters for the spectra with high temperature or low temperature.
The $RRNet_3$ is used to predict the stellar parameters of spectra with low metallicity.
And the $RRNet_4$ is used to improve the parameters of the cold end spectra of dwarf stars.

\subsection{Integration of the estimated results from four sub-models}\label{sec:model:estimation}

In practical application, it is unknown about the sub-region of the parameter space to which a spectrum belongs before estimating its parameters.
This problem makes it impossible to determine which model should be used to predict the spectral parameters beforehand.
To this end, we proposed a strategy of multi-label fusion to solve this problem (Part B in Figure \ref{fig:train_model}).
Based on this strategy, we input a spectrum $\mathbf{X} \in R^{1 \times 7200}$ into $RRNet_{i}, i\in \{1,2,3, 4\}$ respectively. 
The outputs of $RRNet_{i}$  are $\mu_{i}(\mathbf{X}) \in R^{1 \times 16}$ and $\sigma_{i}^{2}(\mathbf{X}) \in R^{1 \times 16}$.
The $\mu_{ij}(\mathbf{X})$ is the estimation of the j-th spectral parameter from $RRNet_{i}$. The $\sigma_{ij}^{2}(\mathbf{X})$ is the uncertainty estimation of $\mu_{ij}(\mathbf{X})$.
The final prediction of the SPEMR model ${\hat\mu}(\mathbf{X})=({\hat\mu_1}(\mathbf{X}),\cdots,{\hat\mu_{17}}(\mathbf{X}))$ and its uncertainty estimate ${\hat\sigma^2}(\mathbf{X})= ({\hat\sigma^2_1}(\mathbf{X}),\dots, {\hat\sigma^2_{17}}(\mathbf{X}))$ can be obtained by fusing \{$\mu_{ij}(\mathbf{X})$, i=1, 2, 3, 4\} and \{$\sigma_{ij}^{2}(\mathbf{X})$, i=1, 2, 3, 4\}.
The specific fusion formula as follows:
\begin{equation}\label{equ2}
   {\hat\mu_j}(\mathbf{X}) = \mu_{i(j)_0,j}(\mathbf{X}),
    \end{equation}
    \begin{equation}\label{equ22}
   {\hat\sigma^2_j}(\mathbf{X}) = {\sigma_{i(j)_0,j}^2}(\mathbf{X}),
\end{equation}
where $i(j)_0 = \arg\min\limits_{i= 1,2,3, 4}{\sigma_{ij}^2}(\mathbf{X})$, and $j=1, \cdots, 16$. That is to say, $i(j)_0$ denotes the model index with the smallest prediction uncertainty ${\sigma_{ij}^2}(\mathbf{X})$.
Therefore, the model fusion schemes (\ref{equ2}) and (\ref{equ22}) are to adopt the predictions with the smallest uncertainty as the final fusion result.

\subsection{Testing of the SPEMR model}\label{sec:test_SPEMR}
\begin{figure*}
	\centering
	\begin{minipage}{1.00\linewidth}
		\centering
		\includegraphics[width=1.0\linewidth]{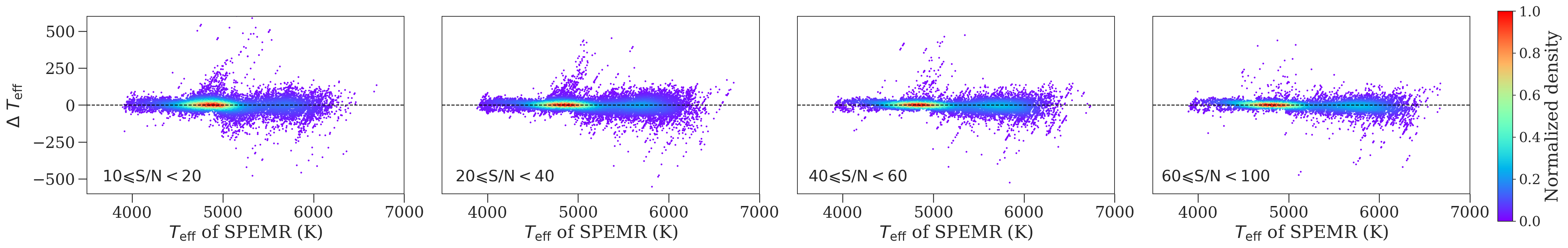}
	\end{minipage}
	\qquad
	\begin{minipage}{1.00\linewidth}
		\centering
		\includegraphics[width=1.0\linewidth]{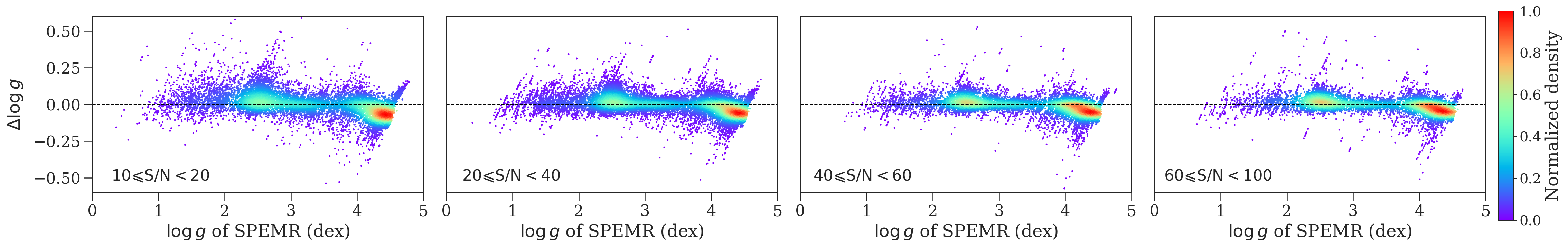}
	\end{minipage}
	\qquad
	\begin{minipage}{1.00\linewidth}
		\centering
		\includegraphics[width=1.0\linewidth]{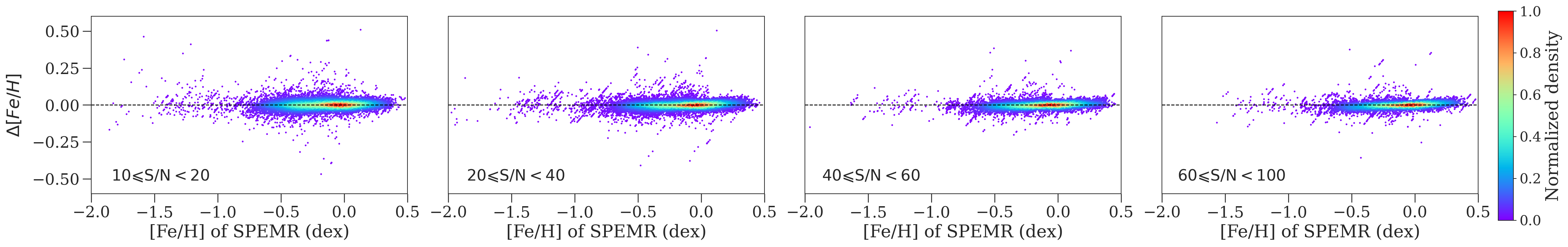}
	\end{minipage}
\caption{Performance evaluation of stellar atmospheric parameters ({$T_{\rm eff}$, $\log \, g$, [Fe/H]}) estimated by SPEMR on spectra with different S/N. The S/N intervals are $\rm 10 \leq S/N < 20$, $\rm 20 \leq S/N < 40$, $\rm 40 \leq S/N < 60$ and $\rm 60 \leq S/N < 100$, respectively. The vertical axis presents the distribution of differences between SPEMR predictions and ASPCAP results on the test set.}
\label{fig:SPEMR_teff_logg_Feh}	
\end{figure*}

\begin{figure*}
\centering
\includegraphics[scale=0.46]{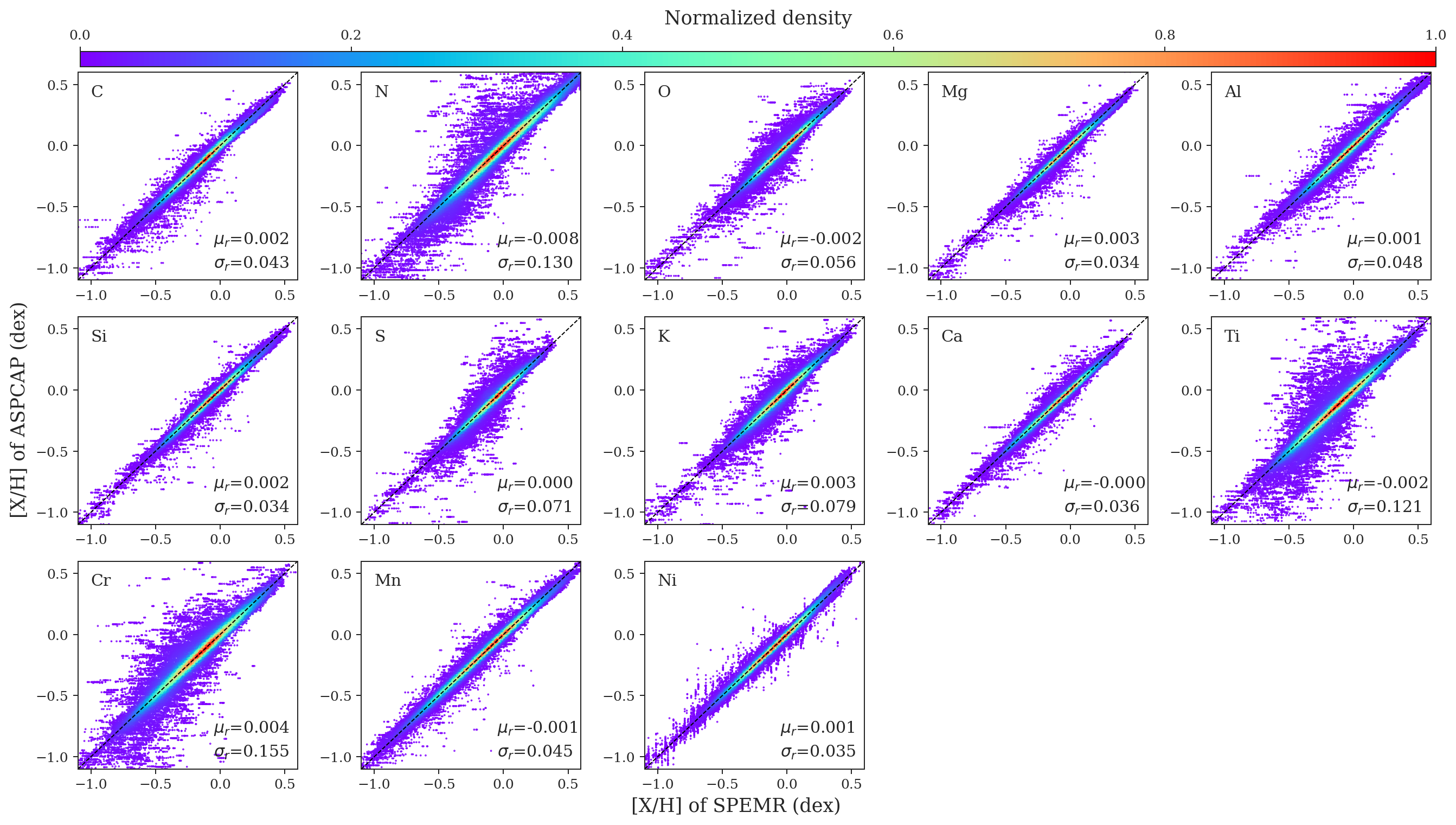}
\caption{Distribution of the residuals between the abundance of 13 elements estimated by SPEMR and the ASPCAP results on the test set. The color indicates the distribution density of the samples.}
\label{fig:SPEMR_Residual_analysis_elemental}
\end{figure*}

After an overall pre-training and three subsequent, independent personalized fine-tuning for the RRNet (Section \ref{subsec:refer_set}), four sub-models are computed: $RRNet_1$, $RRNet_2$, $RRNet_3$ and $RRNet_4$. Based on them, we can obtain the preposed SPEMR model (Section \ref{sec:model:SPEMR}) using the  multi-label fusion strategy (Section \ref{sec:model:estimation}). In this section, we evaluate the performance of the SPEMR model by comparing the differences between the SPEMR estimations and the ASPCAP labels on the test set. Thus, any comparison here is not affected by biases in the ASPCAP results themselves. More comprehensive evaluations are conducted in section \ref{sec:result}.

Figure \ref{fig:SPEMR_teff_logg_Feh} shows the distribution of the differences between the stellar atmospheric parameters predicted by SPEMR and the ASPCAP results.
The deviation of the SPEMR predictions from the ASPCAP labels are small on the spectra with low- and high- S/N level.
This phenonmennon indicates that the SPEMR model can effectively suppress the noise effects on the spectra with low signal-to-noise.
For effective temperature, the corresponding residual is smallest on the spectra with $T_{\rm eff}\in [4500, 5000] $ K.
This phenomenon is mainly due to the large number of training samples in this region and their good quality.
In the second row of Figure \ref{fig:SPEMR_teff_logg_Feh}, we can see a slight underestimation from SPEMR on $\log \, g$ in case of $\log \, g > 4$ (dex).
This phenomenon is consistent with the estimations of \cite{wang2020spcanet} and \cite{xiong2022model}.
It is mainly due to the scarcity of training examples in this parameter space region (as shown in Fig. \ref{fig:refer_dis_Teff_logg}), which causes the increase of prediction error for the dwarfs ($\log \, g > 4$).
For metal abundance, the best prediction results were obtained when the spectra with [Fe/H] $\in$[-0.5, 0.5] dex. This phenomenon is also due to the larger number of training samples in this region.
The above phenomena suggest that the ASPCAP labels provide excellent learning benchmarks for the stellar atmospheric parameters estimated by the SPEMR model.

To further evaluate the results of other parameters estimated by the SPEMR model, we investigated the differences between the SPEMR estimations and the ASPCAP results for the remaining elemental abundances on the test set.
Figure \ref{fig:SPEMR_Residual_analysis_elemental} shows the distribution of differences between the abundances of 13 elements predicted by the SPEMR model and the ASPCAP labels on the test set.
The residual and dispersion of most element abundances estimated by SPEMR are around 0.005~dex and 0.07~dex.
These lower residual and dispersion indicate the better precision and accuracy of the SPEMR model.
However, for elements N, Ti and Cr, the corresponding residual and dispersion are slightly higher.
Therefore, the accuracy of the SPEMR model on elements N, Ti, Cr should be further improved.

\subsection{Best Fitting Template}\label{Best-fit}

\begin{figure*}
\centering
\includegraphics[scale=0.28]{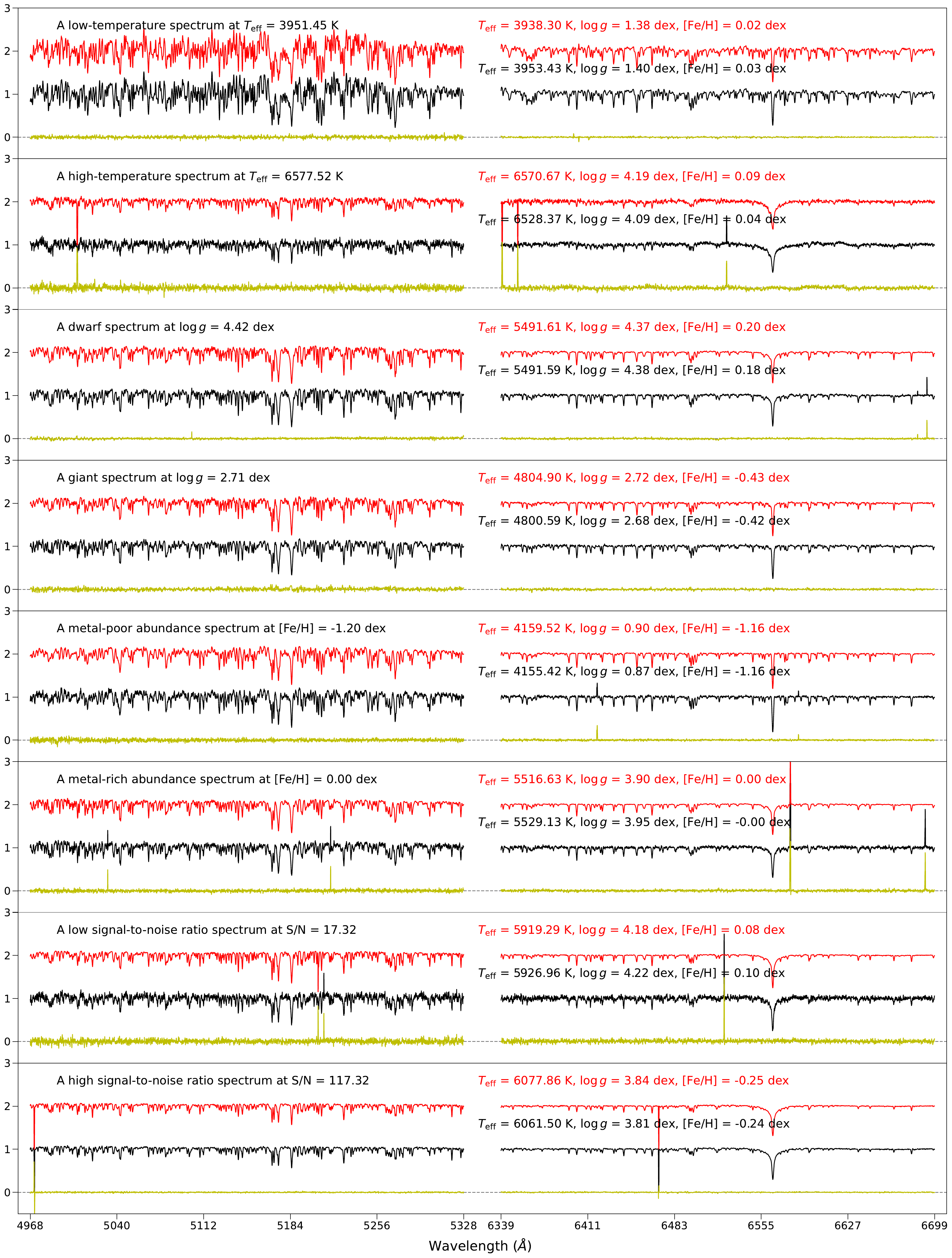}
\caption{Several representative LAMOST spectra with various configurations on parameters and their corresponding best fitting templates. The black line indicates a test spectrum, the red line indicates its best-fit template, and the yellow line indicates the residual between them. The best fit template is determined based on the Euclidean distance between the test spectrum and each training spectrum in stellar parameter space (the parameters of the test spectrum are estimated using the SPEMR model). This figure only shows three stellar atmospheric parameters for each spectrum due to space limitations.}
\label{fig:Best-Fit-Spectrum}
\end{figure*}

To further explore the performance of the SPEMR model, we investigated several representative LAMOST spectra in the test set and their corresponding best-fit templates. These test spectra are selected based on their representativeness in parameter space and spectral quality. This study can increase the physical interpretability of the model and allow the reader to more intuitively observe the fit of the model. In this paper, the best fitting template of a spectrum is found by minimizing the Euclidean distance between the spectrum and each training spectrum using the estimated stellar parameters for test spectra and the reference parameters for the training spectra (the parameters of the test spectrum are estimated using the SPEMR model). The corresponding results are shown in Figure \ref{fig:Best-Fit-Spectrum}. The eight representative LAMOST spectra from top to bottom in Fig. \ref{fig:Best-Fit-Spectrum} are a low-temperature spectrum ($T_{\rm eff}$ = 3951.45 K), a high-temperature spectrum ($T_{\rm eff}$ = 6577.52 K), a dwarf spectrum ($\log \, g$ = 4.42 dex), a giant spectrum ($\log \, g$ = 2.71 dex), a metal-poor abundance spectrum ([Fe/H] = -1.20 dex), a metal-rich abundance spectrum ([Fe/H] = 0.00 dex), a low signal-to-noise ratio spectrum (S/N = 17.32), and a high signal-to-noise ratio spectrum (S/N = 117.32). It can be found that the SPEMR model fits well on low temperature spectrum, dwarf spectrum, giant spectrum, rich metal abundance spectrum, and high signal-to-noise ratio spectrum, that is, the residuals between the test spectrum and the best-fit template almost reach zero in the whole wavelength space. However, for high-temperature spectrum and low-SNR spectrum, the fitting effect of the SPEMR model is not as good as that of the above spectra, especially in the blue-end wavelength space. This phenomenon may be related with the sparse distribution of this kind training samples (Fig. \ref{fig:refer_dis}).

\section{APPLICATION ON LAMOST DR8}
\label{sec:result}

In this section, we applied the SPEMR proposed in section \ref{sec:model} to the LAMOST DR8 medium-resolution spectra to obtain a LAMOST-SPEMR catalog. 
To assess the reliability of the LAMOST-SPEMR catalog, we compared it with other typical catalogs, performed an uncertainty analysis, and tested it on open clusters.

\subsection{Parameter estimation for the medium resolution spectra from LAMOST DR8}

\begin{figure*}
\centering
\includegraphics[scale=0.28]{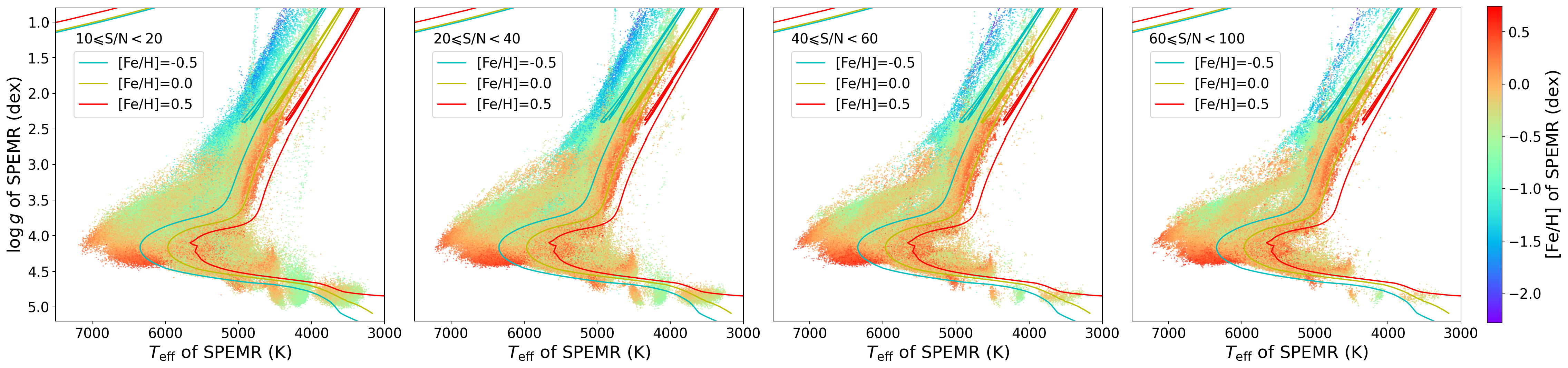}
\caption{Distribution of stellar parameters from LAMOST-SPEMR catalog.
The color in this figure represents $\rm [Fe/H]$, and the three isochrones represent the evolutionary tracks of MIST stars with stellar ages of 7 Gyr ($\rm [Fe/H]$ = -0.5 (cyan), 0.0 (yellow), and 0.5 (red) respectively).
The S/N intervals are $\rm 10 \leq S/N < 20$, $\rm 20 \leq S/N < 40$, $\rm 40 \leq S/N < 60$ and $\rm 60 \leq S/N < 100$, respectively.}
\label{fig:SPEMR_LAMOST_pred_snr}
\end{figure*}

\begin{figure*}
\centering
\includegraphics[scale=0.42]{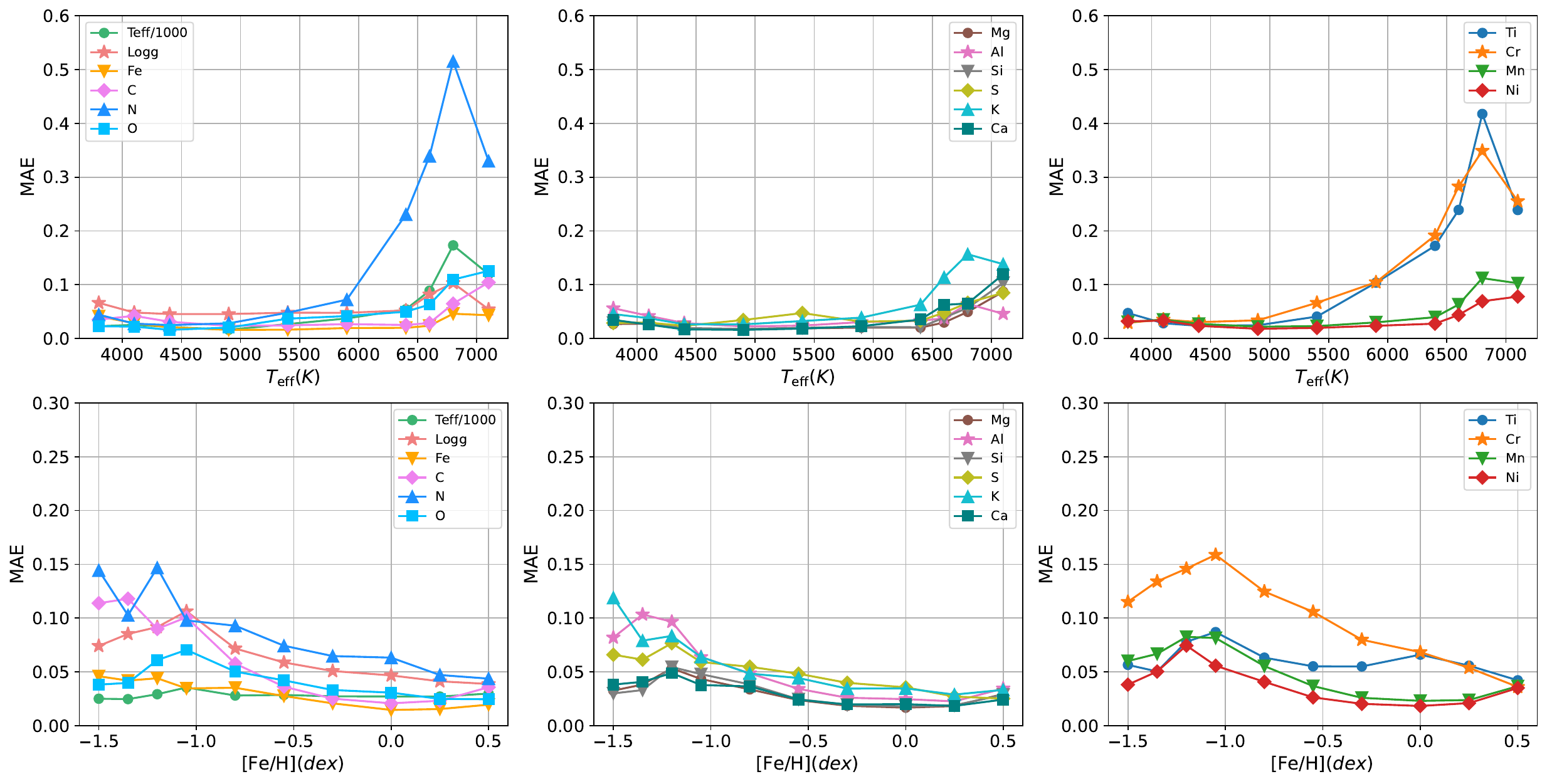}
\caption{The parameter estimation results of SPEMR: the performance on the spectra with low temperature, high temperature, low metallicity, and high-frequency-observed-type spectra are improved to various degrees over the RRNet model (Fig. \ref{fig:Teff_feh}).}
\label{fig:SPEMR_Teff_feh}
\end{figure*}

Three learned sub-models based on reference datasets 1, 2,, 3 and 4($S_1$, $S_2$, $S_3$ in Fig. \ref{fig:refer_dis} and $S_{4}$ in Fig. \ref{fig:refer_dis_Teff_logg}) can perform parameter estimation for spectra on different regions of the parameter space (Section \ref{sec:model:SPEMR}).
The four sub-models can estimate stellar parameters for four types of spectra: the spectra with $T_{\rm eff} \in [4000, 6500]$ K and [Fe/H] $\geq -1.0$ dex, the spectra with $T_{\rm eff}$ > 6500 K or $T_{\rm eff}$ < 4000 K, the spectra with [Fe/H] < -1.0 dex and the spectra with $T_{\rm eff}$ < 5000 K and $\log \, g$ > 2.5 dex, respectively.
The results of four RRNet sub-models are fused using the SPEMR scheme (Section \ref{sec:model:estimation}) to obtain the SPEMR (RRNet) parameter estimation.
Accordingly, the LAMOST-SPEMR catalog is obtianed by SPEMR. This catalog contains stellar atmospheric parameters, chemical abundances, and the corresponding $1\sigma$ uncertainties for 4,197,960 medium-resolution spectra in LAMOST DR8 estimated by SPEMR.

Figure \ref{fig:SPEMR_LAMOST_pred_snr} shows the $T_{\rm eff}-\log \,g$ distribution of the LAMOST-SPEMR catalog in different S/N intervals.
The three isochrones in the figure are the MIST stellar evolutionary tracks with a stellar age of 7 Gyr \citep[][]{dotter2016mesa, choi2016mesa}.
Compared with Figure 9 in \cite{wang2020spcanet}, it is shown that the stellar parameters estimated by SPEMR in the high-temperature spectral region fit better with the three MIST stellar evolutionary tracks. 
In the low-temperature spectral region, the SPEMR and SPCANet estimates show a similar pattern, with an underestimation of $\log \,g$ for the cold ends of the main sequence stars ($T_{\rm eff} \in [4000, 4500]$ K).
For the spectra with low metallicity, SPEMR can also effectively estimate their stellar parameters. 
Compared with Fig. 5 in \cite{xiong2022model}, it is shown taht RRNet lacks the estimation results for the high-temperature spectra, while the proposed SPEMR effectively estimate the stellar parameters from this kind spectra. And the estimation results generally agree with the MIST stellar evolutionary tracks.

To evaluate the validity of SPEMR, we estimated the parameters for the spectra in the reference set, and the results are shown in Fig. \ref{fig:SPEMR_Teff_feh}. 
Compared with the estimation results of RRNet (Fig. \ref{fig:Teff_feh}), the performance of our model is improved to different degrees for the spectra with low temperature, high temperature, low metallicity, and high-frequency-observed-type spectra. 
Specifically, obvious improvements are shown on the spectra with low temperature and the spectra with low metallicity. 
However, no evident improvements are found on the spectra with high temperature.
This phenomenon in the spectra with high temperature is caused by the small number of high-temperature spectral samples in the training data.
In addition, SPEMR improves the estimation results on high-frequency-observed-type spectra, such as $T_{\rm eff}$, $\log \,g$, Fe, Si, etc.

\subsection{Some comparisons with other typical catalogs}

\begin{table*}

\centering
\caption{Some comparisions between the LAMOST-SPEMR catalog and several typical catalogs. The '$\cdots$' indicates that the estimate for a stellar parameter is not given in the corresponding catalog.}
\label{tab1}    
   
\begin{tabular}{lccc||ccc||cc}
    
    \hline
    
     Labels&
     \multicolumn{3}{c}{{SPEMR-ASPCAP}} & 
     \multicolumn{3}{c}{{SPCANet-ASPCAP}} &
     \multicolumn{2}{c}{{SPEMR-GALAH}} \\ 
    
   \cline{2-9}
                  &$\mu_{r}$& $\sigma_{r}$  &MAE      &  $\mu_{r} $ &  $\sigma_{r}$ &MAE  &  $\mu_{r}$  &  $\sigma_{r}$  \\
 \hline
  $T_{\rm eff}$                 &-5.28 	    &57.02 	&33.51  &  -30.65    &118.05    &94.96   &15.03   	&243.65  \\    
  $\log \, g$              &-0.020  	&0.084 	&0.060  &  0.024    & 0.167    &0.118     &0.097  	&0.255   \\
  {[}Fe/H{]}   &-0.005  	&0.038 	&0.026  &  -0.029    &0.075      &0.059   &0.027  	&0.117   \\
  {[}C/H{]}     &-0.009  	&0.063 	&0.045  & -0.115    & 0.119      &0.138   &-0.037  	&0.154   \\
  {[}N/H{]}    &-0.001  	&0.172 	&0.099  &  -0.029    &  0.213      &0.139  &...    &...  \\
  {[}O/H{]}    &-0.005  	&0.083 	&0.054  & -0.035     &  0.129      &0.103  &-0.020  	&0.214   \\
  {[}Mg/H{]}  &-0.005  	&0.046 	&0.033  &  -0.027   &  0.095      &0.075    &0.019  	&0.133   \\
  {[}Al/H{]}    &-0.009  	&0.069 	&0.049  &  -0.051   &  0.157    &0.124  &0.025  	&0.150   \\
  {[}Si/H{]}     &-0.004  	&0.048 	&0.034  &  -0.018   &  0.110    &0.088  &0.032  	&0.110   \\
  {[}S/H{]}      &-0.005  	&0.089 	&0.060  &  -0.019   &  0.116     &0.081  &...    &...  \\
  {[}K/H{]}      &-0.006  	&0.099 	&0.064  &  ...      &  ...        &...                 &-0.014  	&0.258   \\
  {[}Ca/H{]}     &-0.005  	&0.052 	&0.035  &  -0.026   &  0.085     &0.070   &-0.042  	&0.146   \\
  {[}Ti/H{]}     &-0.009  	&0.148 	&0.090  &  0.026   &  0.190      &0.136   &-0.158  	&0.211   \\
  {[}Cr/H{]}     &-0.005  	&0.187 	&0.106  &  0.077    &  0.210      &0.114  &-0.098  	&0.168   \\
  {[}Mn/H{]}     &-0.005  	&0.063 	&0.042  &  ...      &  ...        &...                    &0.017  	&0.149   \\
  {[}Ni/H{]}     &-0.005  	&0.050 	&0.036  &  -0.034    &  0.084      &0.064  &0.025  	&0.141   \\   
    \hline 
\end{tabular}
\begin{flushleft}
NOTE: The {$\mu_r$, $\sigma_r$} are the mean and standard deviation of the difference between two catalogs, respectively.
\end{flushleft}
\end{table*}

\begin{figure*}
\centering
\includegraphics[scale=0.28]{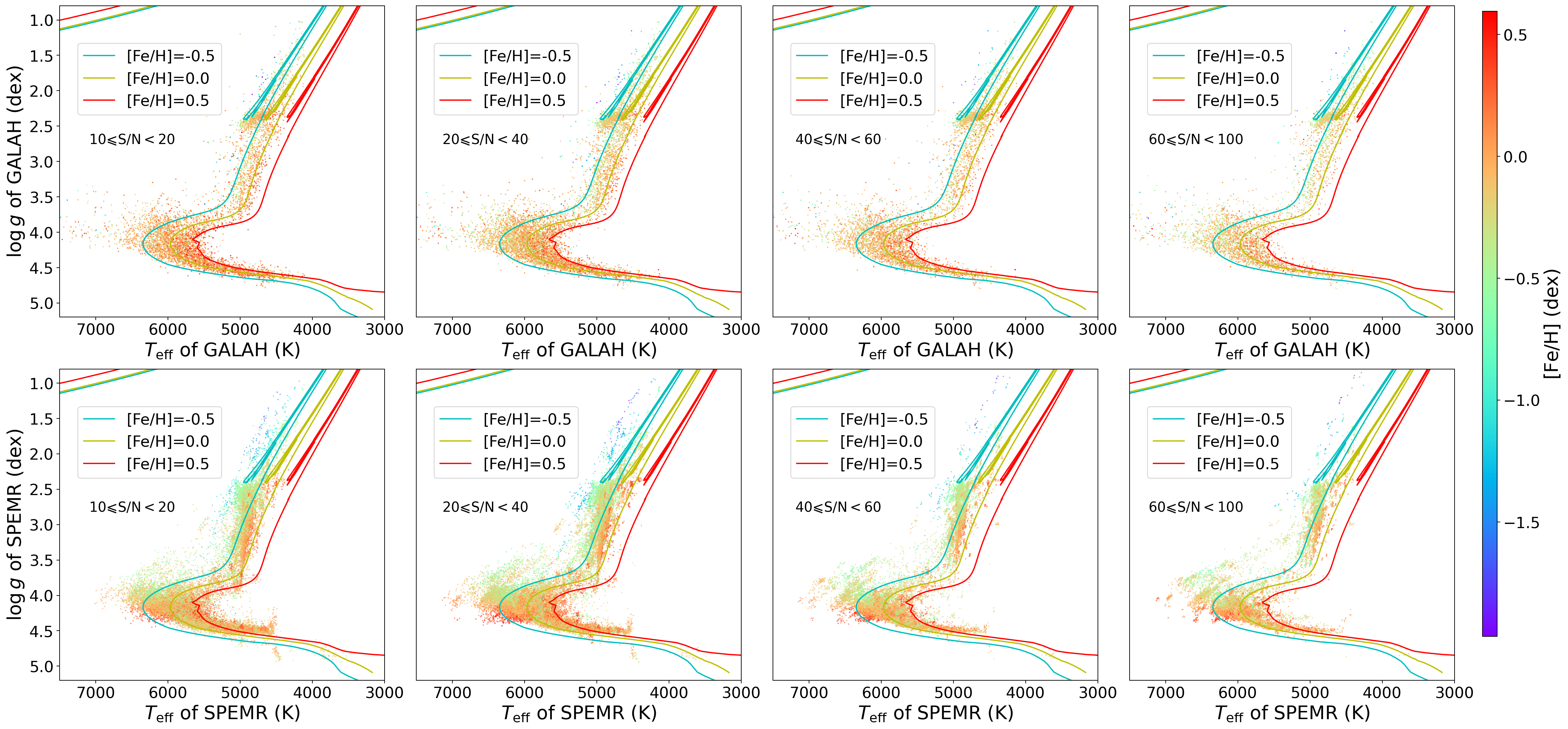}
\caption{Distribution of stellar parameters from GALAH catalog (the first row) and LAMOST-SPEMR catalog (the second row).
The color in this figure represents $\rm [Fe/H]$, and the three isochrones represent the evolutionary tracks of MIST stars with stellar ages of 7 Gyr ($\rm [Fe/H]$ = -0.5 (cyan), 0.0 (yellow), and 0.5 (red) respectively).
The samples in the figure are the 110,042 spectra obtained by cross-matching the LAMOST-SPEMR catalog with the GALAH catalog. The S/N intervals are $\rm 10 \leq S/N < 20$, $\rm 20 \leq S/N < 40$, $\rm 40 \leq S/N < 60$ and $\rm 60 \leq S/N < 100$, respectively.}
\label{fig:GALAH_LAMOST_pred_snr}
\end{figure*}

\begin{figure*}
\centering
\includegraphics[scale=0.45]{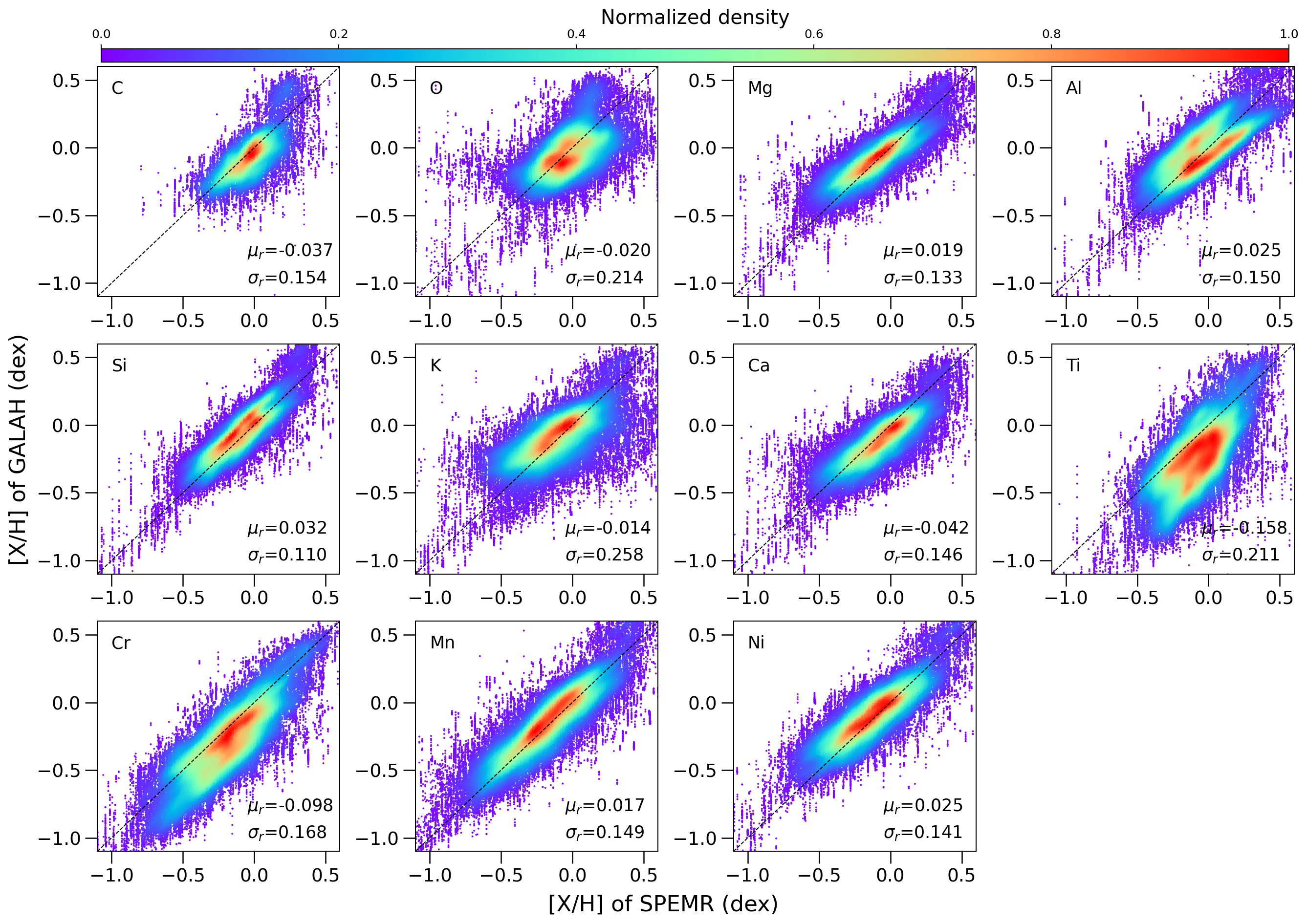}
\caption{Comparison of chemical abundances predicted by LAMOST-SPEMR catalog with GALAH survey’ results. The dotted lines above are theoretical reference lines.}
\label{fig:GALAH_SPEMR_Ref_XH}
\end{figure*}

\begin{figure*}
\centering
\includegraphics[scale=0.45]{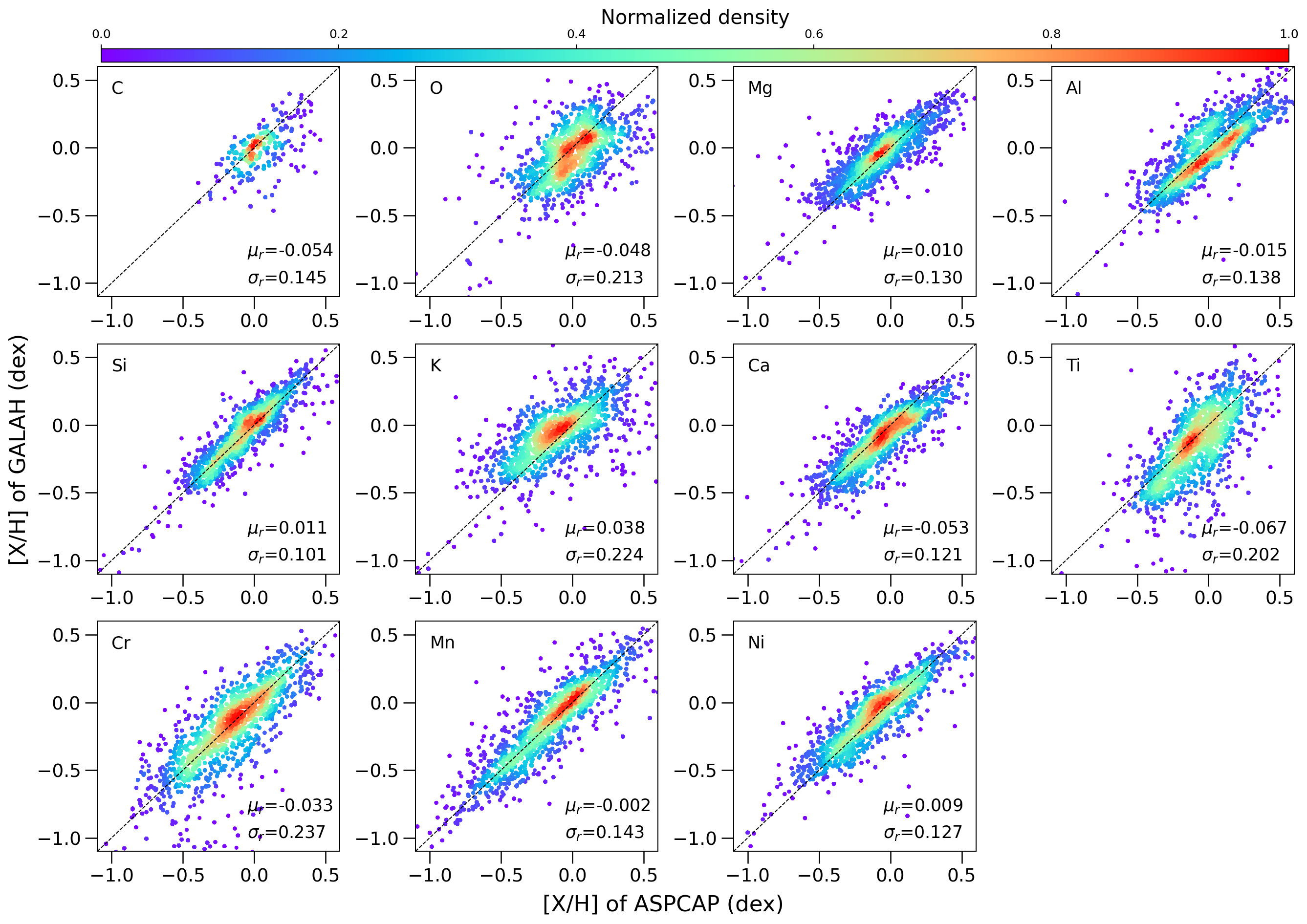}
\caption{Comparison of ASPCAP estimations with GALAH survey’ results on the common sources between APOGEE DR17, LAMOST DR8 medium-resolution spectral library and GALAH catalog. The dotted lines above are theoretical reference lines.}
\label{fig:GALAH_ASPCAP_Ref_XH}
\end{figure*}

\begin{figure*}
    \centering
    \begin{minipage}{0.47\linewidth}
        \centering
        \includegraphics[width=1.0\linewidth]{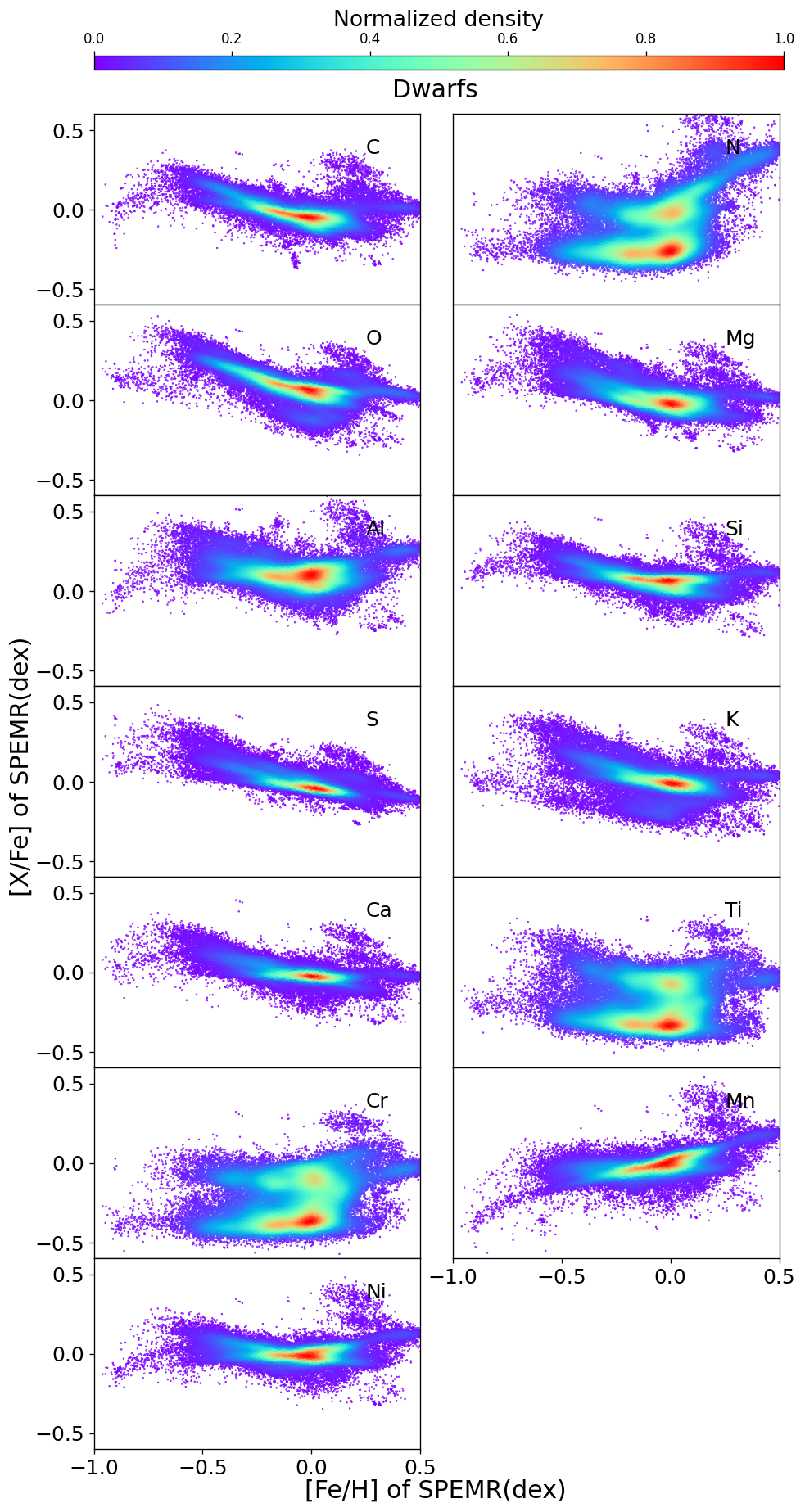}
        \label{fig:SPEMR_LAMOST_FeH_XFe_dwarf}
    \end{minipage}
    \hspace{.02in}
    \begin{minipage}{0.47\linewidth}
        \centering
        \includegraphics[width=1.0\linewidth]{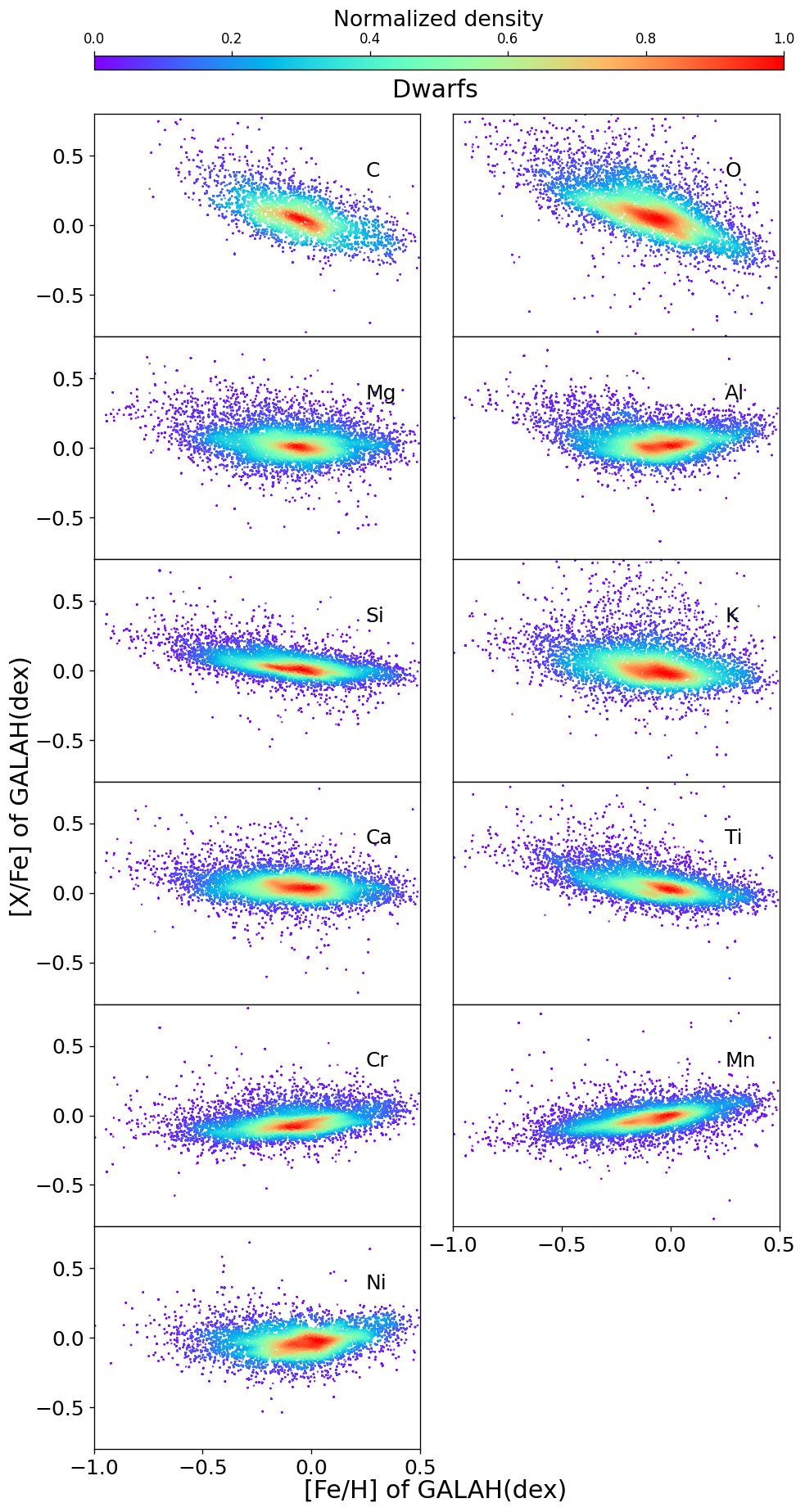}
        \label{fig:APOGEE_FeH_XFe_dwarf}
    \end{minipage}

\caption{Comparison of the distribution density of dwarf ($\log \,g$ > 4) elemental abundances [X/Fe]-[Fe/H] for the LAMOST-SPEMR catalog and GALAH catalog. The two left columns are the estimations for the LAMOST-SPEMR catalog, and the two right columns are the results for the GALAH catalog. The color indicates the density of the sample distribution. Actually, the corresponding samples in the two left columns are from the training set, validation set and test set.}
\label{fig:SPEMR_GALAH_FeH_XFe_dwarf}
\end{figure*}

\begin{figure*}
    \centering
    \begin{minipage}{0.47\linewidth}
        \centering
        \includegraphics[width=1.0\linewidth]{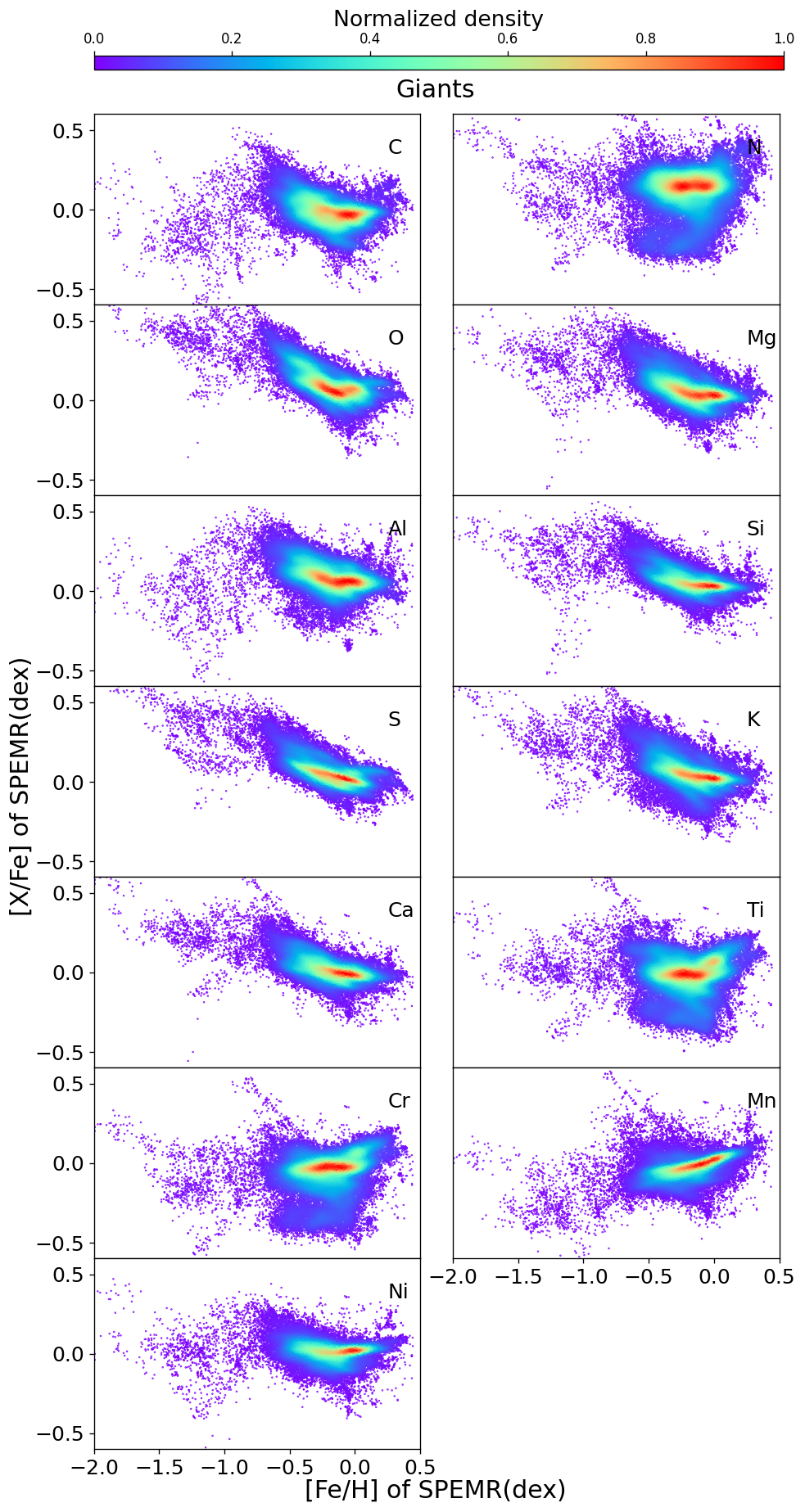}
        \label{fig:SPEMR_LAMOST_FeH_XFe_giant}
    \end{minipage}
    \hspace{.02in}
    \begin{minipage}{0.47\linewidth}
        \centering
        \includegraphics[width=1.0\linewidth]{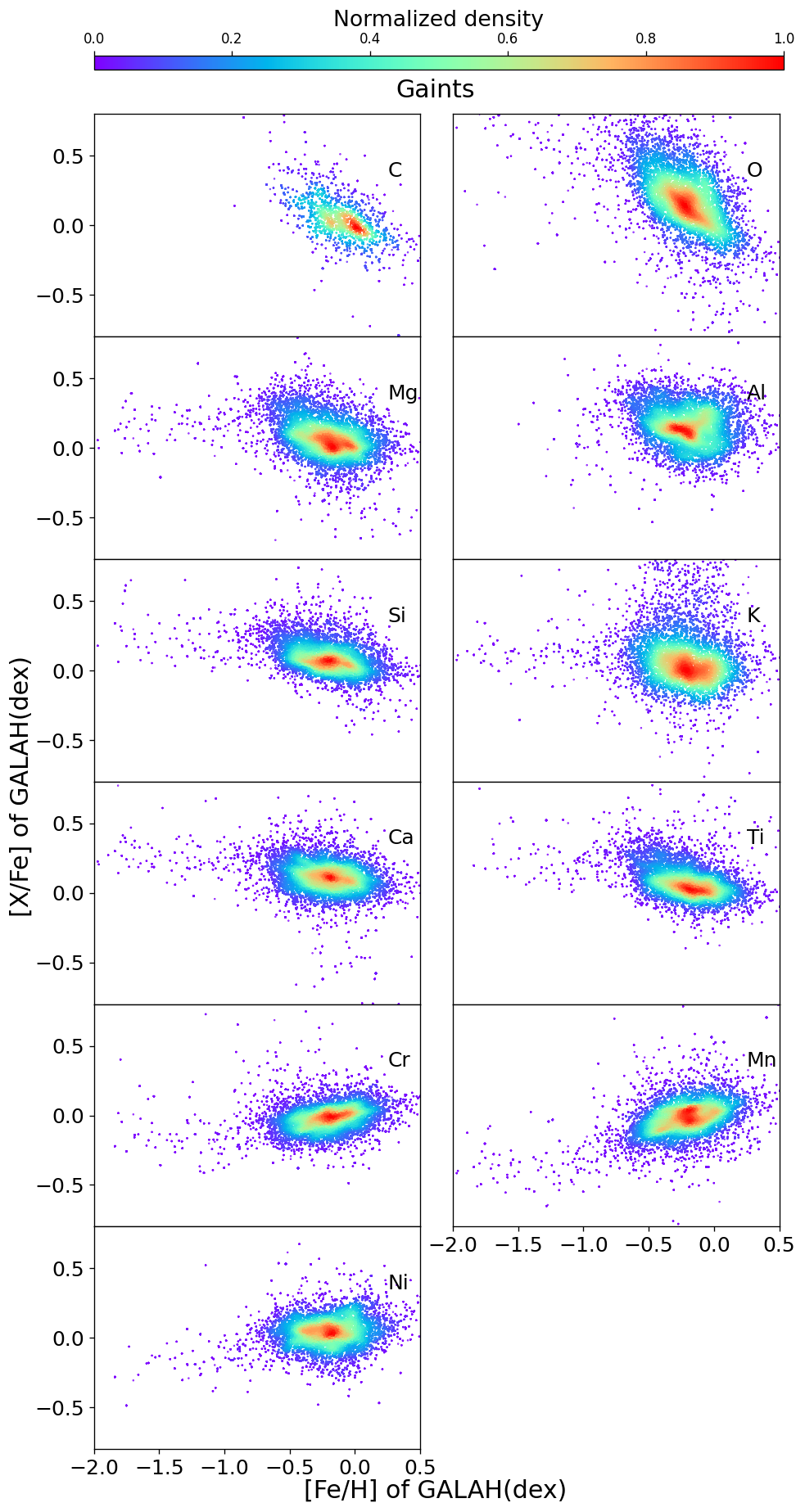}
        \label{fig:APOGEE_FeH_XFe_giant}
    \end{minipage}

\caption{Comparison of the distribution density of giant ($\log \,g$ < 4) elemental abundances [X/Fe]-[Fe/H] for the LAMOST-SPEMR catalog and GALAH catalog. The two left columns are the estimations for the LAMOST-SPEMR catalog, and the two right columns are the results for the GALAH catalog. The color indicates the density of the sample distribution. Actually, the corresponding samples in the two left columns are from the training set, validation set and test set.}
\label{fig:SPEMR_GALAH_FeH_XFe_giant}
\end{figure*}

To further verify the accuracy of the LAMOST-SPEMR catalog, we investigated the consistency between LAMOST-SPEMR catalog and the SPCANet catalog, the GALAH DR3 catalog.

\cite{wang2020spcanet} used the SPCANet model to estimate the stellar atmospheric parameters and chemical abundances for 1,472,211 medium-resolution spectra from LAMSOT DR7, and those results are called the SPCANet catalog in short.
In order to compare the differences between the LAMOST-SPEMR catalog and the SPCANet catalog, we cross-matched the reference set in this paper with the SPCANet catalog and obtained 24,1033 spectra from 5,3775 common stars.
It should be noted that the stellar parameter and chemical abundance estimations are made simutaneously by the SPCANet model, the SPEMR model, and the APOGEE ASPCAP pipeline for each of the 24,1033 spectra.

GALAH \citep{de2015galah} is a large-scale high-resolution (R$\sim$28000) spectroscopic survey project, which uses the Anglo-Australian Telescope and the HERMES spectrograph at the Australian Observatory to observe stellar spectra.
The GALAH spectral coverage is [4713, 4903] \text{\AA}, [5648, 5873] \text{\AA}, [6478, 6737] \text{\AA} and [7585, 7887] \text{\AA}.
GALAH DR3 \citep{buder2021galah+} published stellar atmospheric parameters and elemental abundances for 588,571 stars.
In the observations, there are 383,088 dwarf stars, 200,927 giant stars, and 4,556 unclassified stars.
We cross-matched the LAMOST-SPEMR catalog with the GALAH DR3 catalog and obtained 110,042 LAMOST DR8 spectra from 25,519 common stars. 

Compared with the SPCANet catalog (Table \ref{tab1} (SPEMR-ASPCAP, SPEMR-ASPCAP)), LAMOST-SPEMR estimated two more parameters [K/H] and [Mn/H], and reduced the overall bias, dispersion, and MAE by 30\%, 50\%, 52\%, respectively on most of the stellar parameters. 
This phenomenon indicates that SPEMR has better estimation performance than SPCANet.
However, for elements S, Ti, Cr, the precision improvement of the SPEMR model are not significant.
This may be caused by the lack of stronger metal lines in the blue part of the LAMOST spectra.

To further investigate the consistency between the LAMOST-SPEMR catalog and GALAH DR3 catalog, we evaluated the differences between SPEMR and GALAH results. Figure \ref{fig:GALAH_LAMOST_pred_snr} shows the $T_{\rm eff}-\log \,g$ distribution of the GALAH catalog and LAMOST-SPEMR catalog colored by [Fe/H] at different S/N intervals. 
Compared with the GALAH catalog, the LAMOST-SPEMR catalog shows a larger dispersion for giant stars with effective temperature around 5000 K, especially in the region of low metallicity ([Fe/H] < -0.5 dex). This phenomenon is mainly due to the scarcity of reference samples with ($T_\mathrm{eff}$, log $g$) $\sim$ (5000 K, 2.5 dex) (Figure \ref{fig:refer_dis_Teff_logg}), as well as the relatively sparse spectral samples for [Fe/H] < -0.5 dex (Figure \ref{fig:APOGEE_DR17_distribution} and Figure \ref{fig:refer_dis}). In addition,  the LAMOST-SPEMR catalog shows some close correlations between the fitting performance to the MIST stellar evolution tracks and the signal-to-noise ratio (SNR). In case of a higher SNR, the LAMOST-SPEMR catalog shows a stronger consistency with MIST stellar evolution tracks, and the corresponding dispersion and bias are also smaller. Otherwise, the corresponding dispersion and bias are larger on the low signal-to-noise spectra. This is mainly due to the poor quality of the low SNR spectra and it indicates that it is necessary to investigate a more robust estimation method and establish an expanded reference set with a better coverage on stellar parameter space. It is worth mentioning that \cite{li2022estimating} have specifically studied the parameter estimation problem from the LAMOST stellar spectra with low SNR and low resolution. Therefore, we can also pay special attention to the low SNR LAMOST medium-resolution spectral data to improve the overall parameter estimation accuracy of the model in future research.

Figure \ref{fig:GALAH_SPEMR_Ref_XH} shows the comparison of the abundances of chemical elements estimated by LAMOST-SPEMR catalog with the results of the GALAH catalog. The detailed biases and the standard deviations are listed in Table \ref{tab1} (SPEMR-GALAH). It is shown that the standard deviations of the difference between the LAMOST-SPEMR catalog and the GALAH catalog range between 0.13 dex $\sim$ 0.15 dex in the abundance of elements C, Mg, Al, Si, Ca, Mn, and Ni, and the overall differences distribute around the theoretical line with little dispersion. For the elements O, K, Ti, and Cr, the corresponding differences have a relatively larger dispersion, around 0.20 dex; and the estimations of Ti and Cr show some relatively evident deviation from the theoretical line. To further explore the source of the discrepancy between the LAMOST-SPEMR catalog and the GALAH catalog, we compared the ASPCAP reference value with those of the GALAH catalog in terms of the above elemental abundances (Figure \ref{fig:GALAH_ASPCAP_Ref_XH}). It is shown that the trend of the difference between the ASPCAP catalog and the GALAH catalog is basically consistent with that of the LAMOST-SPEMR catalog. This indicates that the deviation and dispersion of the difference between LAMSOT-SPEMR results and GALAH survey largely originate from the difference between the ASPCAP catalog and the GALAH catalog.

To further evaluate the performance of the SPEMR model, we show the [X/Fe] vs. [Fe/H] distributions of all elements estimated by SPEMR for giants and dwarfs and compare them to the GALAH results for the same stars. Figure \ref{fig:SPEMR_GALAH_FeH_XFe_dwarf} shows the [X/Fe]-[Fe/H] distribution over the dwarfs ($\log \,g > 4$ dex) for all elements of the LAMOST-SPEMR and GALAH DR3 catalogs. Figure \ref{fig:SPEMR_GALAH_FeH_XFe_giant} shows the distribution of the corresponding giants ($\log \,g < 4$ dex). Comparing the two left and right columns in Fig. \ref{fig:SPEMR_GALAH_FeH_XFe_dwarf} and Fig. \ref{fig:SPEMR_GALAH_FeH_XFe_giant}, we can clearly find that the elemental abundance patterns of LAMOST-SPEMR catalog are tighter than those of the GALAH DR3 catalog both on the dwarfs and on the giants. For the dwarfs, the elemental abundances of the LAMOST-SPEMR catalog are concentrated in the intermediate metal abundances ([Fe/H] $\in$ [-0.2, 0.3]dex).
For the giants, the distribution of elemental abundances is much wider, and most of them are concentrated in [Fe/H] $\in$[-0.6, 0.4]dex. This phenomenon is largely consistent with the distribution of the GALAH DR3 catalog.
Comparing the two left columns of Fig. \ref{fig:SPEMR_GALAH_FeH_XFe_dwarf} and Fig. \ref{fig:SPEMR_GALAH_FeH_XFe_giant}, we can find that the distributions of giants and dwarfs are inconsistent for most elements estimated by SPEMR. For example, the elements Mg, Al, Ti, and Cr from the giants show a more dense distribution on [Fe/H] from -0.8 dex to 0.3 dex. However, this dense pattern is not present in the dwarfs. This is mainly due to the scarcity of main-sequence dwarfs in our training samples. This sample imbalance leads to the fact that most of the labels predicted by SPEMR are around red giants. Only the elements Cr and Ti of the dwarfs show distinct bimodal structures, while most of the elements of the giants all show distinct bimodal structures. 
In addition, for the giants, elements O, S and K show clear negative correlations relative to [Fe/H], while elements N, Cr and Mn show obvious positive correlations relative to [Fe/H]. 
And the other elements are closely distributed on a horizontal line. 
For the dwarfs, elements C, O, and S show  obvious negative correlations relative to [Fe/H], while elements N, AL, Cr and Mn show obvious positive correlations relative to [Fe/H]. 
And other elements are closely distributed on a horizontal line. 
For most elemental abundances, the position and slope of the dwarf stars distributions are not consistent with those of the giant stars.This phenomenon may be caused by the difference on sampling spaces of the dwarf and giant stars. 

\subsection{Uncertainty Analysis}
\label{sec:Uncertainty}

\begin{figure*}
\centering
\includegraphics[scale=0.42]{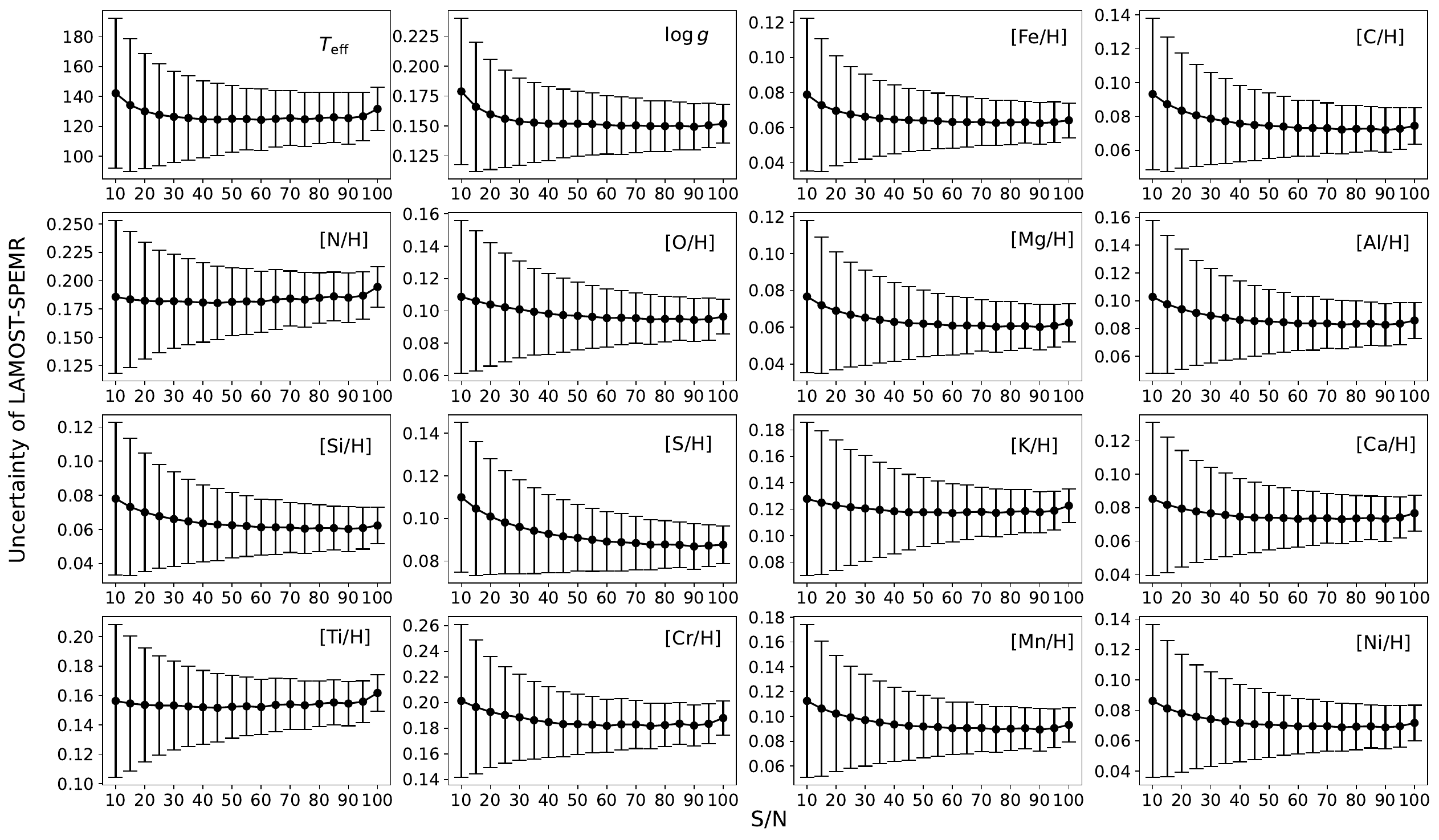}
\caption{Dependencies of parameter estimation uncertainties on S/N. 
The dots in this figure indicate the uncertainty predicted by SPEMR, and the length of the line segments centered on the dots indicate the uncertainty obtained from repeated observations (> 5 times).}
\label{fig:uncertainties_analysis_snr}
\end{figure*}

The SPEMR model is able to predict the PDF of stellar parameters and gives the uncertainty $\sigma_{pred}$ for the predicted parameters using a deep ensembling approach and equation (\ref{equ:final_variance}).
In addition, in the LAMOST sky survey, some stars are observed for multiple times at different time and under various observing conditions.
This phenomenon can be used to analyze the uncertainty $\sigma_{obs}$ caused by observation configurations.
Suppose we have $n_s$ repeated observations $\{\mathbf{x}_1, \cdots, \mathbf{x}_{n_s} \}$ from a source and $n_s$ estimations $\{y_1, \cdots, y_{n_s} \}$ from these observations using SPEMR.
The standard deviation of these $n_s$ parameter estimations is the corresponding uncertainty $\sigma_{obs}$.

Figure \ref{fig:uncertainties_analysis_snr} shows the dependencies of the uncertainty of LAMOST-SPEMR catalog on S/N.
The dots in this figure indicate the uncertainties $\sigma_{pred}$ predicted by SPEMR, and the length of the line segment centered on the dots indicate the uncertainty estimated from the repeated observations $\sigma_{obs}$.
On the whole, the RRNet model shows the strong robustness and generalization from the lower uncertainty.
Specifically, in case of S/N $\geq 10$, the $\sigma_{pred}$ of the parameters $T_{\rm eff}$, $\log \,g$ and [Fe/H] are 134 K, 0.17 dex and 0.07 dex, respectively, and those of the remaining elements are 0.07 dex$\sim$0.19 dex.
In addition, $\sigma_{pred}$ and $\sigma_{obs}$  decrease with the increase of S/N.
This phenomenon is caused by the higher quality of LAMOST spectra with high S/N. 
The spectra with high S/N suffer from less noises. 
The uncertainties in this paper are numerically different from Figure 7 in \cite{xiong2022model}. This difference is caused by the region changes in the parameter space of stellar spectra under being processed.
At the same time, RRNet and SPEMR show a similar pattern. Therefore, the SPEMR model is stable in estimating stellar parameters from LAMOST spectra.

\subsection{Tests on open clusters}

\begin{figure*}
\centering
\includegraphics[scale=0.40]{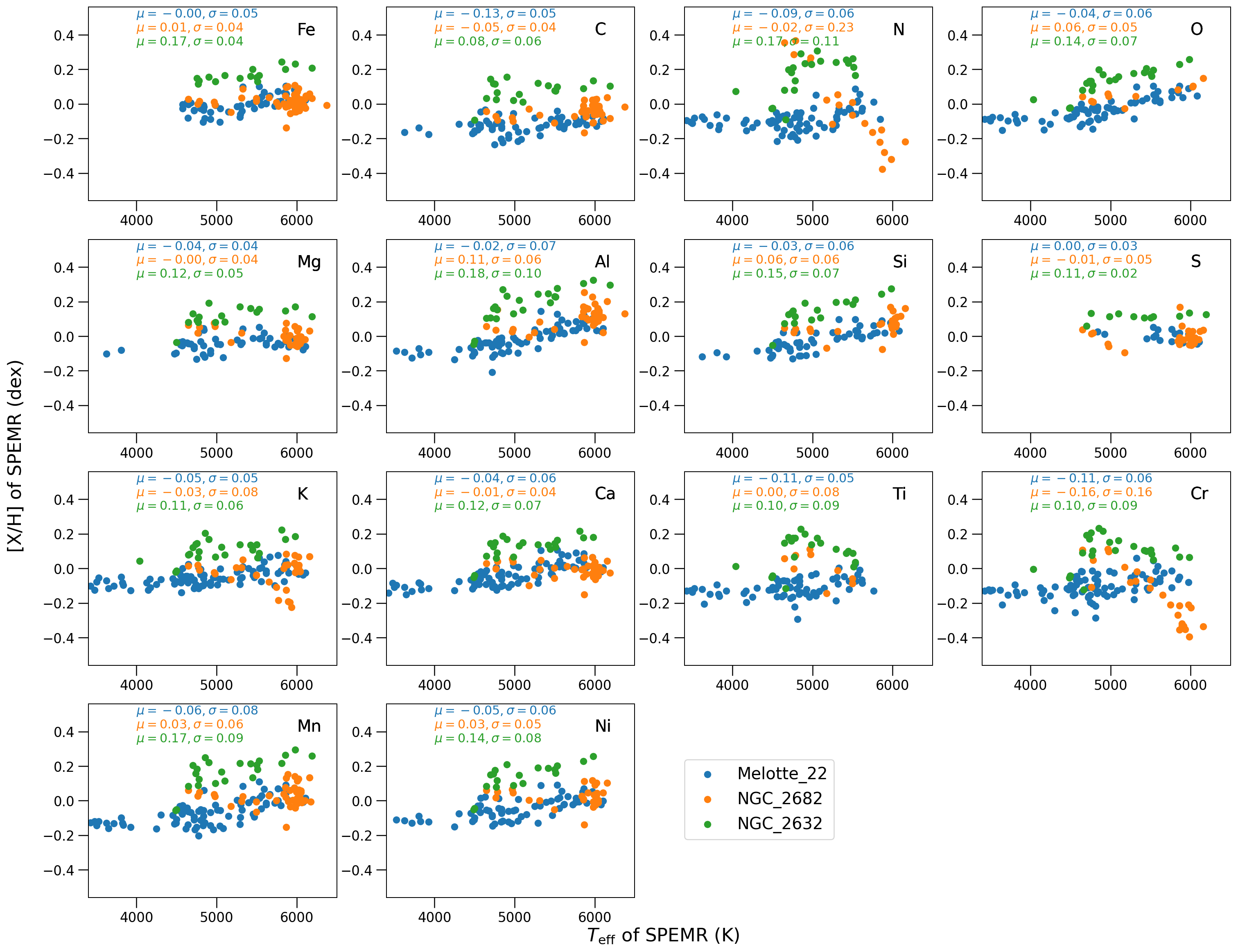}
\caption{The variation of chemical abundances from the LAMOST-SPEMR catalog with $T_{\rm eff}$ in the three open clusters of Melotte 22, NGC 2682, and NGC 2632.
The three colors in the figure correspond to the three open clusters, and the mean $\mu$ and standard deviation $\sigma$ of the chemical abundances are added to each panel.}
\label{fig:open_clusters_analysis_Teff}
\end{figure*}

\begin{figure*}
\centering
\includegraphics[scale=0.40]{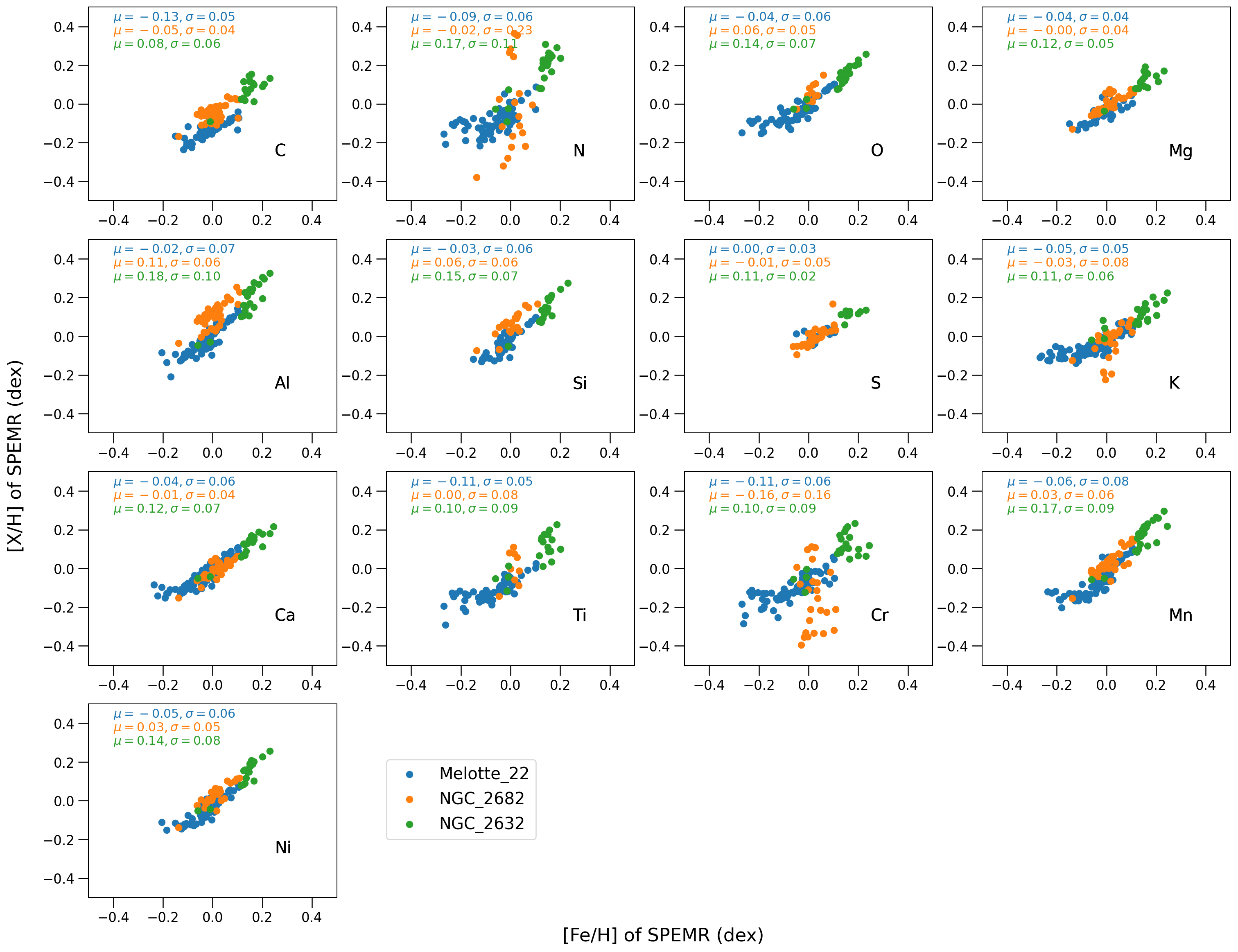}
\caption{The variation of chemical abundances from the LAMOST-SPEMR catalog with [Fe/H] for the three open clusters of Melotte 22, NGC 2682, and NGC 2632. The three colors in the figure correspond to the three open clusters, and the mean $\mu$ and standard deviation $\sigma$ of the chemical abundances are added to each panel.}
\label{fig:open_clusters_analysis_Feh}
\end{figure*}

\begin{figure*}
\centering
\includegraphics[scale=0.65]{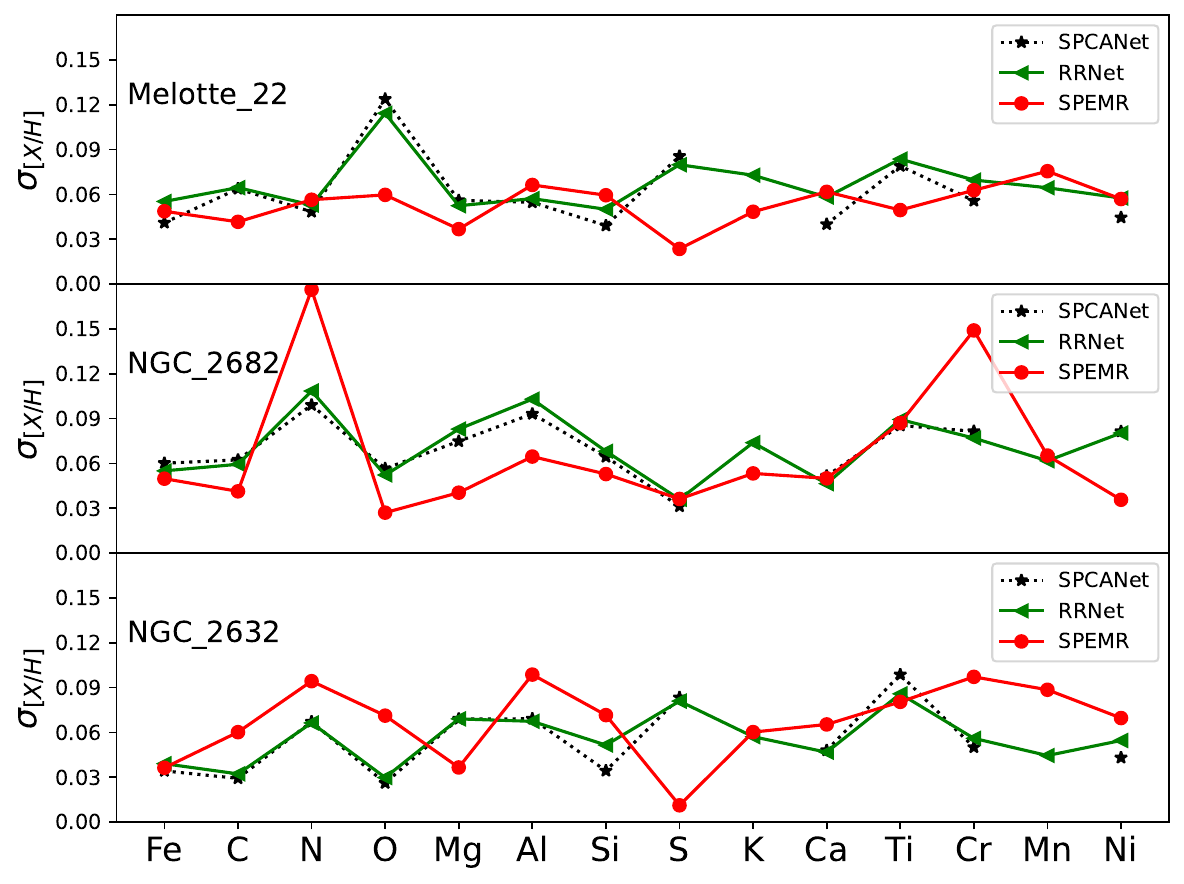}
\caption{Some comparisons of LAMOST-SPEMR with the SPCANet and RRNet catalogs in terms of chemical abundances in three open clusters (Melotte 22, NGC 2682, NGC 2632).}
\label{fig:open_clusters_analysis_comp_SPCANet}
\end{figure*}

Open clusters have good chemical homogeneity \citep[][]{bovy2016chemical, Ness2018Galactic}.
Therefore, they can be used as chemical indicators to assess the effect of stellar parameter estimation.
To further investigate the accuracy of the element abundances from the LAMOST-SPEMR catalog, we performed more tests on open clusters.
\cite {Zhong2020Exploring} analyzed the properties of many open clusters based on Gaia DR2 and LAMOST data, and provided a spectroscopic parametric catalog consisting of the stellar physical parameters of 8,811 member stars.
We cross-matched these cluster member stars with the LAMOST-SPEMR catalog, and obtained a variety of open clusters, such as Melotte 22, NGC 2682, NGC 2632, NGC 2168, Melotte 20, NGC 2281, Stock 2, NGC 1750, NGC 1545, and so on. 
Finaly, we selected three open clusters (Melotte 22, NGC 2682 and NGC 2632) with the largest number of matches to LAMOST-SPEMR and removed the parameter estimation with large uncertainties $\sigma_{pred}$ (section \ref{sec:Uncertainty}). 

To investigate the effects of effective temperature and metal abundance on the elemental abundances from open clusters, we show the variation of LAMOST-SPEMR chemical elemental abundances with effective temperature and metal abundance in the three open clusters.
Figure \ref{fig:open_clusters_analysis_Teff} shows the dependencies of the chemical elemental abundances from LAMOST-SPEMR catalog on effective temperature ($T_{\rm eff}$) in the above-mentioned open clusters.
In agreement with the performance of \cite{Ting_2019}, the chemical abundances of SPEMR do not show a significant variation trend with $T_{\rm eff}$ in all three aforementioned clusters, and the chemical abundances are at a low deviation and dispersion.
Figure \ref{fig:open_clusters_analysis_Feh} shows the dependencies of the chemical elemental abundances from LAMOST-SPEMR catalog on metal abundance ([Fe/H]) for the above-mentioned open clusters. It is shown that [Fe/H] values estimated by the LAMOST-SPEMR catalog approximately range  between -0.2 dex and 0.2 dex. And there is a different [Fe/H] spread depending on [X/H] for these open clusters. This phenomenon is mainly due to the differences between these open clusters in ages and distances \citep{GaiaESO2015}.

In addition, we compared the chemical abundances of the LAMOST-SPEMR catalog with those of the SPCANet and RRNet catalogs on the three above-mentioned clusters (Fig. \ref{fig:open_clusters_analysis_comp_SPCANet}).
Figure \ref{fig:open_clusters_analysis_comp_SPCANet} shows the standard deviation of the elemental abundances from LAMOST-SPEMR and RRNet catalogs are lower than those of the SPCANet catalog on the whole.
For the Melotte 22 and NGC 2632 open clusters, LAMOST-SPEMR shows an overall lower standard deviation than RRNet. 
For the NGC 2682 open cluster, LAMOST-SPEMR shows lower standard deviation than RRNet on elements (S, Ca, Ti, and Cr). 
The overall chemical homogeneity from LAMOST-SPEMR on the three clusters is {0.054$\pm$0.022 dex}, {0.055$\pm$0.016 dex} and {0.067$\pm$0.024 dex}, respectively. These phenomena indicate that the LAMOST-SPEMR catalog has higher accuracy compared with SPCANet and RRNet.

\subsection{LAMOST-SPEMR catalog}
Finally, we published the LAMOST-SPEMR catalog for the estimated stellar atmospheric parameters and elemental abundances of 4,197,960 medium-resolution spectra from LAMOST DR8.
This catalog contains the following information: the identifier for the observed spectrum (obsid), the fits file name corresponding to the spectrum (filename), coordinate information (ra, dec), the extension name of the spectrum (extname\_blue, extname\_red), the signal-to-noise ratio of the spectrum (snr\_blue, snr\_red), effective temperature (Teff[K]), surface gravity (Logg), metallicity (Fe/H), 13 elemental abundances (X/H), and the $1\sigma$ uncertainty of the corresponding stellar parameters (X\_err).

\section{Summary and outlook}
\label{sec:conclution}

This paper proposed a novel method Stellar Parameter Estimation based on Multiple Regions Scheme (SPEMR)  based on the distribution characteristics of LAMOST medium-resolution data in parameter space.
We estimated the stellar atmospheric parameters, elemental abundances, and corresponding uncertainties for 4,197,960 medium-resolution spectra in LAMOST DR8 using SPEMR.
In case of S/N $\geq 10$, the precision of the parameters $T_{\rm eff}$, $\log \,g$, [Fe/H], and [Cr/H] are 47 K, 0.08 dex, 0.03 dex, and 0.16 dex, respectively, while the precision of the other elemental abundances are 0.03 dex $\sim$ 0.13 dex.
To verify the performance of SPEMR, we conducted a series of comparing experiments with other typical medium-resolution spectral parameter estimation models and other surveys.
The experimental results demonstrate that the SPEMR model not only improves the parameter accuracy on high-frequency-observed-type spectra but also provides good parameter estimation on the spectra with high temperature, low temperature, or low metallicity.
In addition, the SPEMR parameter estimation results are excellently consistent with other high-resolution sky survey. 
In the future, we will explore the characteristics of high-temperature and low-signal-to-noise spectra, and build extended reference sets with a better coverage on stellar parameter space to further improve the parameter estimation capability of the model.

\section*{Acknowledgements}
We are very grateful to the referee for helpful suggestions, as well as the correction for some issues, which have improved the paper significantly.
This work is supported by the National Natural Science Foundation of China (Grant No. 11973022), the Natural Science Foundation of Guangdong Province (No. 2020A1515010710), the Major projects of the joint fund of Guangdong, and the National Natural Science Foundation (Grant No. U1811464).

LAMOST, a multi-target optical fiber spectroscopic telescope in the large sky area, is a major national engineering project built by the Chinese Academy of Sciences. Funding for the project is provided by the National Development and Reform Commission. LAMOST is operated and managed by the National Astronomical Observatory of the Chinese Academy of Sciences.

\section*{Data Availability}

The LAMOST data employed in this article are available after September
2022 to the users out of China for download from LAMOST DR8, at \url{http://www.lamost.org/dr8/}. The computed catalog for 4.19 million medium-resolution spectra from the LAMOST DR8, the source code, the trained model and the experimental data have been made publicly available at: \url{https://github.com/yulongzh/SPEMR}.

\section*{Footnotes}

software: Numpy \citep{harris2020array}, Scipy \citep{virtanen2020scipy}, Astropy \citep{price2018astropy}, Matplotlib \citep{hunter2007matplotlib}, Scikit-learn\citep{pedregosa2011scikit},
Pytorch\citep{paszke2019pytorch}.



\bibliographystyle{mnras}
\bibliography{example} 




\bsp	
\label{lastpage}
\end{document}